\documentclass[11pt,a4paper]{article}
\usepackage{amsmath}
\usepackage[dvipdfmx,hiresbb]{graphicx}
\usepackage{xcolor}
\usepackage{mathrsfs,mathtools}
\usepackage{physics,amssymb}
\usepackage{physics}
\usepackage{siunitx}
\usepackage{bm}
\usepackage{braket}
\usepackage{listings}
\DeclareMathOperator{\sinc}{sinc}
\usepackage{cases}
\usepackage{comment}
\usepackage{soul}
\usepackage{cancel}
\usepackage{cases}
\usepackage[utf8]{inputenc}
\usepackage{url}
\usepackage{float}
\usepackage{longtable}
\usepackage[normalem]{ulem}
\usepackage{xspace}
\usepackage{aas_macros}
\usepackage{subcaption}
\usepackage{autobreak}

\usepackage{ulem}

\usepackage{subcaption}
\usepackage{empheq}
\usepackage{here}
\usepackage{pdfpages}

\setstcolor{red}
\usepackage{jcappub}
\hypersetup{colorlinks=true
,urlcolor=DARKBLUE
,anchorcolor=DARKBLUE
,citecolor=DARKBLUE
,filecolor=DARKBLUE
,linkcolor=DARKBLUE
,menucolor=DARKBLUE
,linktocpage=true
,pdfproducer=medialab
}

\newcommand{\uII}{\mathrm{II}}

\newcommand{\uI}{\mathrm{I}}

\definecolor{MONZA}{HTML}{CF000F}
\definecolor{DARKBLUE}{HTML}{00008b}
 \definecolor{DARKBLUE}{rgb}{0,0,0.7} 
\definecolor{DARKMAGENTA}{HTML}{8b008b}
\definecolor{DARKCYAN}{HTML}{008B8B}
\definecolor{DARKORANGE}{HTML}{FF8C00}
\definecolor{GREEN}{HTML}{008000}

\newcommand{\mathblue}[1]{\textcolor{DARKBLUE}{#1}}

\definecolor{MAGENTA}{HTML}{FF00FF}

\begin{document}

\title{Primordial Black Hole Formation from Type II Fluctuations with Primordial Non-Gaussianity}

\author[a]{Masaaki Shimada,}
\author[a]{Albert Escriv\'{a},}
\author[a]{Daiki Saito,}
\author[a]{Koichiro Uehara,}
\author[a]{and Chul-Moon Yoo}

\emailAdd{shimada.massaaki.h8@s.mail.nagoya-u.ac.jp}
\emailAdd{escriva.manas.albert.y0@a.mail.nagoya-u.ac.jp}
\emailAdd{saito.daiki.g3@s.mail.nagoya-u.ac.jp}
\emailAdd{uehara.koichiro.p8@s.mail.nagoya-u.ac.jp}
\emailAdd{yoo.chulmoon.k6@f.mail.nagoya-u.ac.jp}

\affiliation[a]{Division of Science, Graduate School of Science, Nagoya University, Nagoya 464-8602, Japan}

\date{\today}
\abstract{
This study investigates the formation of primordial black holes (PBHs) resulting from the collapse of adiabatic fluctuations with large amplitudes and non-Gaussianity. 
Ref. \cite{Uehara:2024yyp} showed that fluctuations with large amplitudes lead to the formation of type B PBHs, 
characterized by the existence of the bifurcating trapping horizons,
distinct from the more common type A PBHs without a bifurcating trapping horizon. 
We focus on the local type non-Gaussianity characterized by the curvature perturbation $\zeta$ given by 
a function of a Gaussian random variable $\zeta_{\rm G}$ as $\beta\zeta=-\ln(1-\beta \zeta_{\rm G})$ with a parameter $\beta$. 
Then we examine how the non-Gaussianity influences the dynamics and the type of PBH formed, particularly focusing on type II fluctuations, where the areal radius varies non-monotonically with the coordinate radius. 
Our findings indicate that, for $\beta>-2$,
the threshold for distinguishing between type A and type B PBHs decreases with increasing 
$\beta$ similarly to the threshold for black hole formation. 
Additionally, for large positive values of $\beta$, the threshold for type B PBHs approaches that for type II fluctuations.
We also find that, 
for a sufficiently large negative value of $\beta\lesssim-4.0$, the threshold value is in the type II region of $\mu$, i.e., there are fluctuations of type II that do not form black holes. 
Lastly, we calculate the PBH mass for several values of $\beta$. 
Then we observe that the final mass monotonically increases with the initial amplitude within the parameter region of type A PBHs, which differs from previous analytical expectations. 
}
\maketitle
\flushbottom

\section{Introduction}

Primordial Black Holes (PBHs) are Black Holes (BHs) formed in the early universe, 
which can play a crucial role in exploring the early universe \cite{Zeldovich:1967lct,Carr,Hawking,1980AZh....57..250N}.
They are unique probes of statistical properties of the small-scale primordial inhomogeneity. 
In addition, they are considered as one of the candidates for dark matter and could be detected as black hole binaries by gravitational wave interferometers \cite{Khlopov:2008qy,Sasaki:2018dmp,Carr:2020gox,Green:2020jor,Escriva:2022duf}.
In the standard scenario of PBH formation, PBHs are formed during the radiation-dominated era when regions of extremely high density, generated by fluctuations during the inflationary period, undergo gravitational collapse \cite{Hawking}.
\par
In this scenario, an enhancement of the power spectrum of the curvature fluctuations is typically necessary to produce a considerable fraction of PBHs in the form of dark matter, which can be followed by the generation of non-Gaussianity. Non-Gaussianities can have a significant effect on the PBH formation process and abundance \cite{Bullock:1996at,PinaAvelino:2005rm,Hidalgo:2007vk,Pattison:2017mbe,Atal:2018neu,Cai:2018dkf,Passaglia:2018ixg,Yoo:2019pma,Escriva:2022pnz,Yoo:2022mzl}, and on induced gravitational waves \cite{Cai:2018dig,Domenech:2021ztg,Adshead:2021hnm,Abe:2022xur,Li:2023xtl} (see \cite{Pi:2024jwt} for a focused review on non-Gaussianities).

On the other side, when the initial amplitude of the primordial fluctuation is extremely large, 
we may have a non-monotonic behaviour of the areal radius as a function of the radial coordinate $r$, 
and we call it type II fluctuation following Ref.~\cite{Kopp}, 
while standard fluctuations with monotonic behavior of the areal radius type I fluctuation.
The study of PBH formation associated with type II fluctuations has been conducted using the Lemaitre-Tolman-Bondi (LTB) solution, 
which is the analytic solution for the dust fluid, in Ref.~\cite{Kopp}.
Type II fluctuations may have a very low probability of generation due to their huge amplitude far above its threshold, and they are usually considered to contribute negligibly to the abundance of PBHs.

However, it has been pointed out that if the statistics of the curvature fluctuations exhibit non-Gaussianity, 
the threshold of the amplitude for PBH formation may be close to that for the realization of type II fluctuations, and type II fluctuations may significantly contribute to the PBH abundance~\cite{Nong}.
Moreover, in Ref.~\cite{bubble}, type II fluctuations have been observed surrounding the bubbles (which eventually will produce PBHs \cite{Garriga:2015fdk}) formed when the inflaton overshoots the barrier separating a local minimum of the inflaton potential in a single-field inflationary model. 
As non-Gaussianity increases, the vacuum bubble formation channel becomes dominant in comparison with the adiabatic one. 
As a result, the mass and abundance of PBHs are primarily given by those of type II fluctuations.

Recently, Ref.~\cite{Uehara:2024yyp} simulated PBH formation from type II fluctuations. 
It was proposed to classify PBH types into type A/B depending on the existence of the bifurcating trapping horizons, independently of the type of the initial fluctuations, type I/II.
We also follow this convention in this paper.  
\par

In this work, we mainly explore the behaviour of fluctuations of type II under the effect of non-gaussianities.
In particular, based on the typical profile depending on a non-Gaussian parameter, we investigate the threshold between type A/B PBH and how the PBH mass depends on the initial amplitude of the fluctuation.
Our study aims to give insights and get a deeper understanding of the behavior of fluctuations of type II, 
specifically focusing on the scenario of non-Gaussianities.
This paper is organized as follows.
In Section \ref{peak}, we introduce spherically symmetric curvature fluctuations with non-Gaussianity on the super-horizon scale.
In Section \ref{compaction}, we introduce the compaction function, which serves as an indicator of the threshold for PBH formation, and present the threshold values derived from analytical approaches.
Section \mathblue{\ref{Num}}
presents the numerical simulation setup and the results obtained.
Finally, 
Section \ref{conclusion} is devoted to summary.

Throughout this paper, we use the geometrized units in which both 
the speed of light and Newton's gravitational constant are set to unity, $c=G=1$.

\section{Peak profile with non-Gaussianity}\label{peak}
In \cite{Atal_2020,nong2}, it was shown that in inflationary models with a bump in the potential, the non-Gaussian curvature perturbation $\zeta$ is related to the Gaussian field $\zeta_G$ in a local form, derived by using the $\delta N$ formalism \cite{Sugiyama:2012tj} as 
\begin{equation}
    \zeta = -\frac{1}{\beta}\ln(1-\beta \zeta_G),  \label{eq:zeta}
\end{equation}
where $\zeta_{G}$ is the Gaussian curvature perturbation.%
\footnote{Expanding the expression $\zeta$ up to the quadratic order of $\zeta_G$, 
we find that the non-Gaussianity parameter $f_{\rm NL}$ is related to $\beta$ as $\beta = 6 f_{\rm NL}/5$
at the quadratic order.}
The Gaussian limit can be realized by taking $\beta\rightarrow0$. 
The parameter $\beta$ can be written as a function of the potential as
\begin{equation}
    \beta = \frac{1}{2}\left(-3 + \sqrt{9-12 \frac{V^{\prime\prime}}{V}}\right), 
\end{equation}
where $V$ is the inflationary potential.\footnote{Similar expressions to Eq.~\eqref{eq:zeta} were recently discussed in \cite{Pi:2022ysn}, where more general potentials are considered.}
In the model with a bump in the potential, the second derivative of the potential $V''$ in the relevant region likely to be 
negative, and we expect $\beta\geq0$. 
Nevertheless, we just naively extend the region of $\beta$ to negative values
without introducing any specific model that realizes negative values of $\beta$. 
For $\beta<0$, the probability distribution of $\zeta$ 
is given by the Gumbel like-tail instead of an exponential tail for $\beta>0$, as discussed in \cite{Pi:2022ysn}.

Note that the curvature fluctuation is not well-defined when $\beta\zeta_G> 1$. 
In the specific setting discussed in \cite{Atal_2020,nong2}, 
this forbidden parameter region represents the alternative channel for PBH production made by vacuum bubbles.
Such a scenario has been studied in detail in \cite{bubble}, and we refer the reader for more details. 
In this work, we concentrate on the standard adiabatic channel of PBH production with $\beta\zeta_G<1$.

\begin{figure}[t]
    \centering
    \begin{minipage}[t]{0.45\linewidth}
        \includegraphics[width = \linewidth]{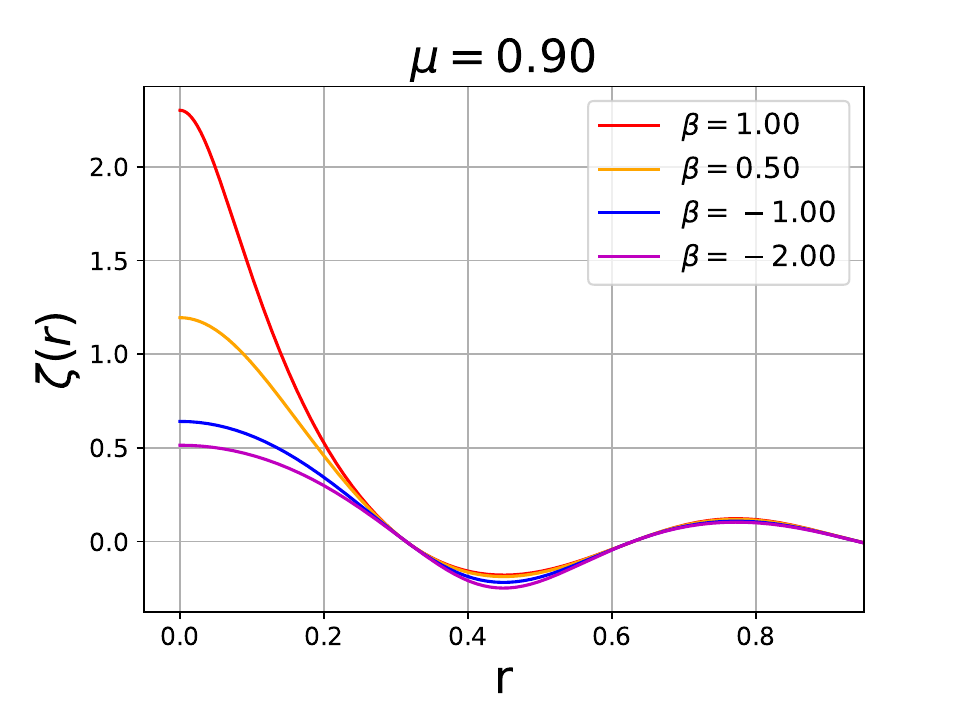}
    \end{minipage}
    \begin{minipage}[t]{0.45\linewidth}
        \includegraphics[width = \linewidth]{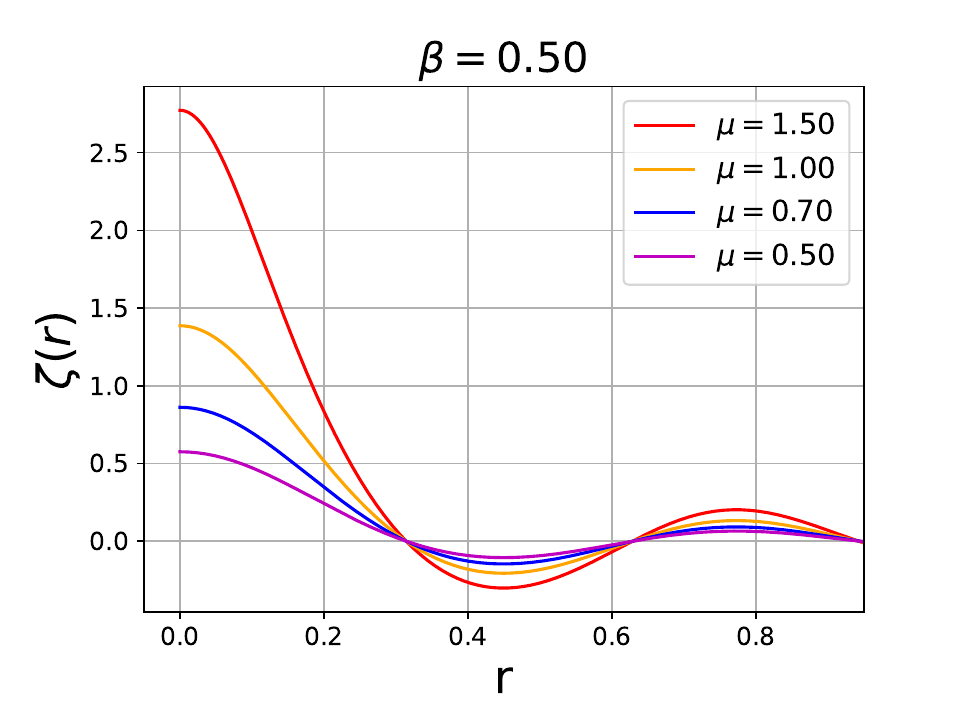}
    \end{minipage}
    \centering
    \begin{minipage}[t]{0.45\linewidth}
        \includegraphics[width = \linewidth]{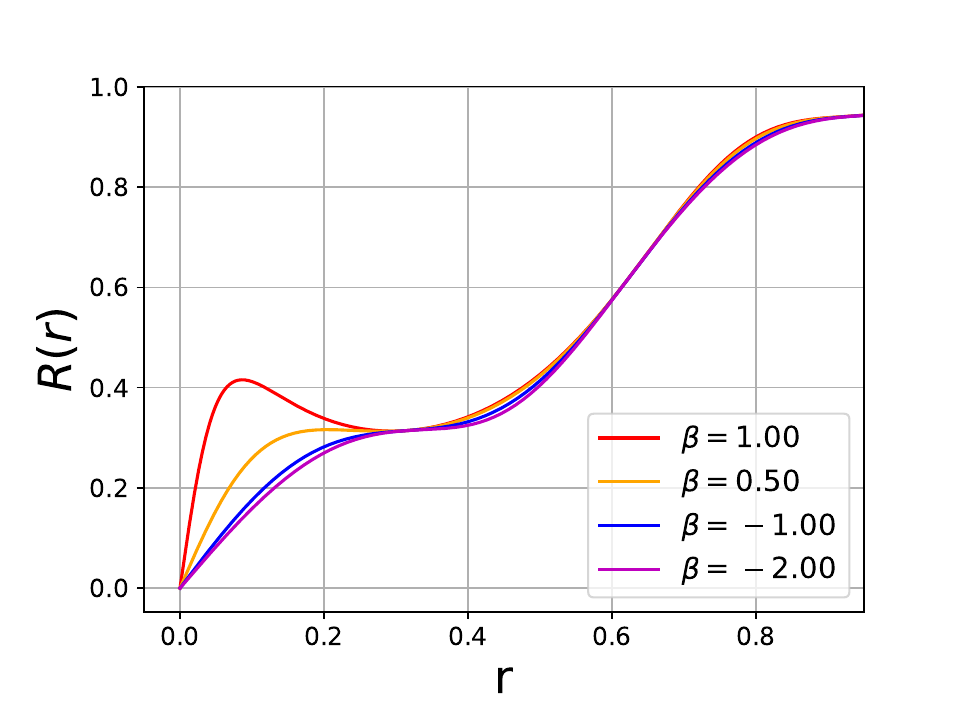}
    \end{minipage}
    \begin{minipage}[t]{0.45\linewidth}
        \includegraphics[width = \linewidth]{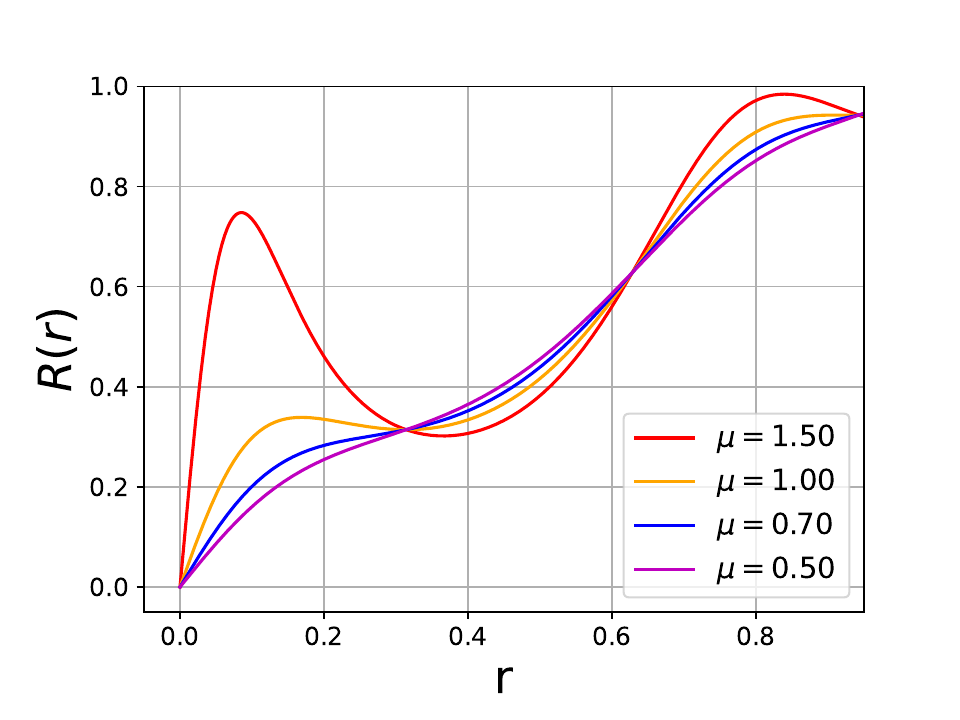}  
    \end{minipage}
    \caption{The profile of $\zeta(r)$ (upper-pannels) and $R(r)$ (lower-pannels) for fixed value of $\mu = 0.90$ (left) and $\beta = 0.50$ (right). The blue and magenta lines represent type $\uI$ fluctuation, and the red and orange lines represent type $\uII$ fluctuation.}\label{fig:Rr}
\end{figure}

We assume Gaussian curvature perturbations sourced by an idealized monochromatic power spectrum\footnote{In principle, the Gaussian curvature fluctuation $\zeta_G$ is determined by solving numerically the Mukhanov-Sasaki equation \cite{10.1143/PTP.70.394,1988ZhETF..94....1M}, which depends on the inflationary model considered.
Thus, in general, the power spectrum should be fixed depending on the inflationary model. 
Nevertheless, for simplicity, we fix $\zeta_G$ to be monochromatic for all $\beta$'s in this work.}. 
Then the typical spherically symmetric profile of $\zeta_G$ can be given by 
$\zeta_{G} = \mu\sinc(kr)$~\cite{1986ApJ...304...15B}, and we obtain
\begin{equation}
\zeta = -\frac{1}{\beta} \ln(1-\beta\mu\sinc(kr)). 
\label{eq:zeta_log}
\end{equation} 
We show the profiles of $\zeta(r)$ in the upper panels of Fig.~\ref{fig:Rr}.
We set initial conditions when these curvature fluctuations are on super-horizon scales, 
meaning that the length scale of the fluctuations is much larger than the Hubble length $H^{-1}_{b}$.
That is $k^{-1} \gg (aH_{b})^{-1}$ with $k, a= a(t)$ and $H_{b} \equiv (da/dt)/a$ being the characteristic comoving wave number, the scale factor and the Hubble parameter of the background flat FLRW universe, respectively.
By applying the cosmological long-wavelength approximation \cite{Harada:2015yda}, 
we can show that, at the leading order of the long-wavelength approximation, $\zeta$ is time-independent.
Then, for a given spatial function $\zeta$,
we can determine the initial conditions up through the next-leading terms at the order of $\epsilon^2$
with $\epsilon = k/(aH_{b})$.
In the leading order, the spherically symmetric metric can be written as follows \cite{Shibata:1999zs}: 
\begin{equation}
    ^{(3)}ds^2  = a^{2}(t)e^{2\zeta(r)}(dr^2 + r^2d\Omega^2),
\end{equation}
where $d\Omega^2$ represents the metric of a 2-sphere.
The areal radius is given by $R=are^{\zeta (r)}$, and we show the profiles of $R(r)$ in the lower panels of Fig.~\ref{fig:Rr}.
Fig.~\ref{fig:Rr} shows that, for a fixed value of $\beta$, 
taking a sufficiently large value of $\mu$ makes $R(r)$ non-monotonic along radial coordinate $r$, i.e., there is the radius $r_{\uII}$ such that $\partial_{r}R(r_{\uII}) =0$ for a sufficiently large value of $\mu$.
We define the fluctuation that satisfies this condition as type II fluctuation and conversely define the fluctuation with a monotonic areal radius as type I fluctuation.
In both the left and right panels of Fig.~\ref{fig:Rr}, 
the red line represents type II fluctuations, while the blue and magenta lines represent type $\uI$ fluctuations.

Now, let us check the condition for type II fluctuation analytically from Eq.~\eqref{eq:zeta}.
The condition can be translated to
\begin{equation}    
1+\left(r\partial_{r}\zeta(r)\right)_{r=r_\uII} = 0.
\label{eq:typeII}
\end{equation}
It should be noted that, for a sufficiently small value of $\mu$, 
there are no roots of $r_{\uII}$. 
From this condition, we can derive the function 
$\mu_\uII(r_{\uII};\beta)$ for which the type II condition \eqref{eq:typeII} is realized at the radius $r=r_\uII$ as follows:
\begin{equation}
    \mu=\mu_\uII(r_{\uII};\beta) = \frac{kr_{\uII}}{(1+\beta)\sin(kr_{\uII}) - kr_{\uII}\cos(kr_{\uII})}.
    \label{eq:mu22}
\end{equation}
Since there are no roots of $r_{\uII}$ for sufficiently small value of $\mu$, 
there is the smallest possible value of $\mu$
which allows the existence of a root. 
The root of $r_{\uII}$ for the smallest possible value of $\mu$, which we denote as $r_{\uI/\uII}$, is a degenerate one, that is, 
$\partial_{r_{\uII}}\mu_{\uII}(r_{\uI/\uII};\beta) = 0$ is satisfied. 
Then we obtain
\begin{equation}
    \beta = \frac{kr_{\uI/\uII}\cos(kr_{\uI/\uII}) - (1-k^2r_{\uI/\uII}^2)\sin(kr_{\uI/\uII})}{\sin(kr_{\uI/\uII}) - kr_{\uI/\uII}\cos(kr_{\uI/\uII})}. \label{eq:betarII}
\end{equation}
Inversely solving \eqref{eq:betarII} for $r_{\uI/\uII}$ as $r_{\uI/\uII}(\beta)$ and 
substituting it into $r_\uII$ in Eq.~\eqref{eq:mu22}, 
we obtain the value of $\mu$ dividing the parameter region into type I and II as $\mu=\mu_\uII(r_{\uI/\uII}(\beta),\beta)$. 
The function $\mu=\mu_\uII(r_{\uI/\uII}(\beta),\beta)$ is depicted in Fig.~\ref{fig:mucbeta} as a solid blue line.

\section{Compaction function and analytical criterion}
\label{compaction}
\begin{figure}[t]
    \centering
    \begin{minipage}[h]{0.32\linewidth}
        \includegraphics[width = \linewidth]{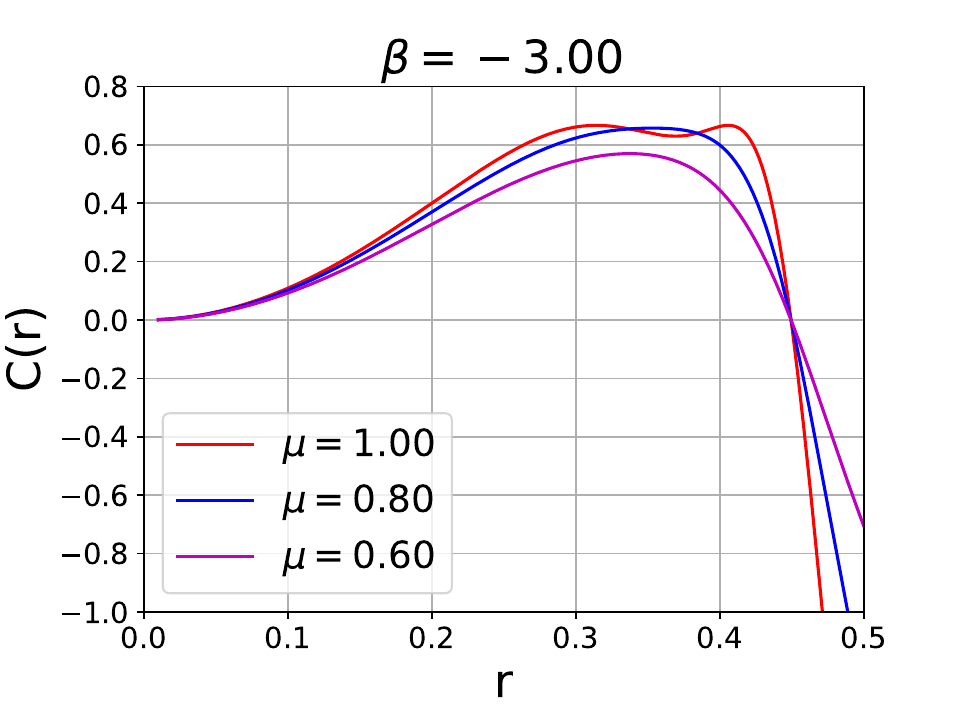}
    \end{minipage}
    \begin{minipage}[h]{0.32\linewidth}
        \includegraphics[width = \linewidth]{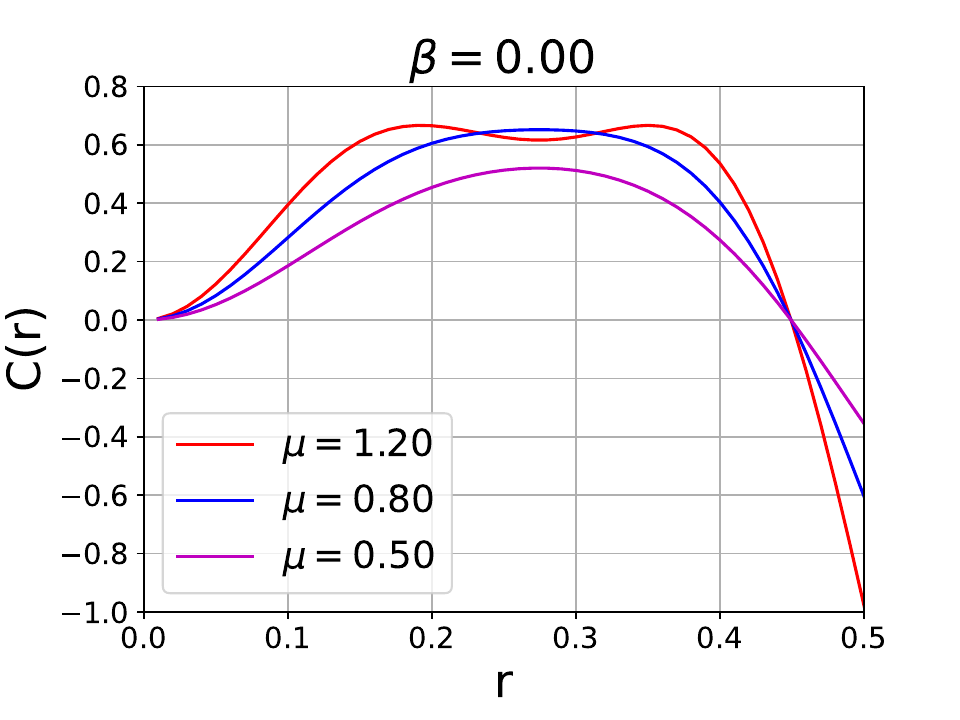}
    \end{minipage}
    \begin{minipage}[h]{0.32\linewidth}
        \includegraphics[width = \linewidth]{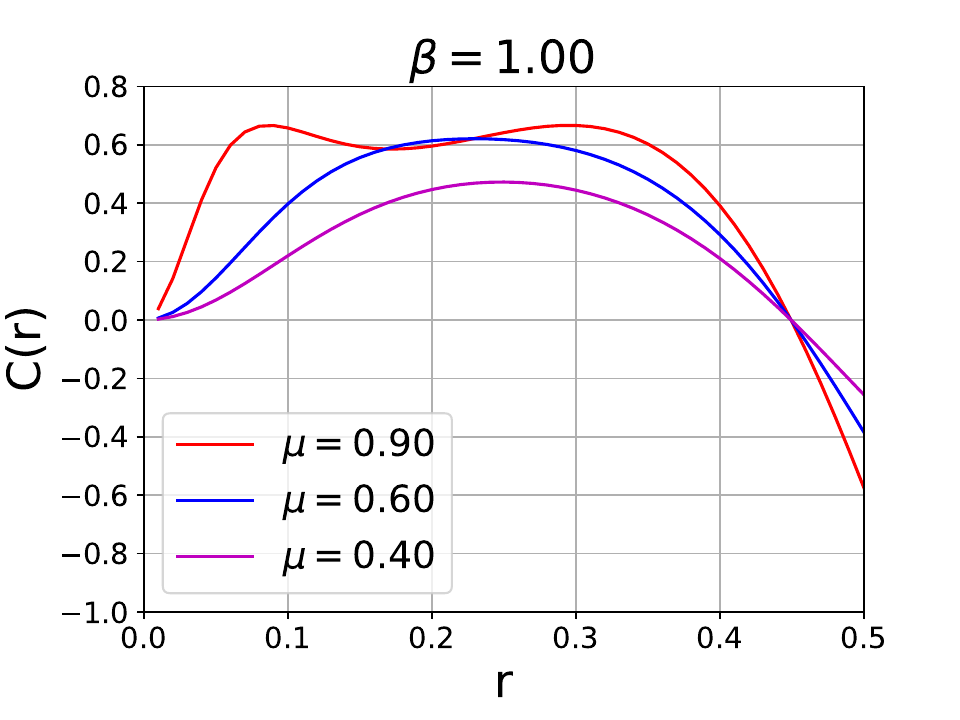}
    \end{minipage}
    \caption{
    The profile of the initial compaction function $\mathcal{C}(r)$ using Eq.~\eqref{eq:compaction} for three $\beta$ cases.
    The blue and magenta curves are for type I fluctuations, and the red curves, which have three extremal points, are for type II fluctuations.} 
    \label{fig:compaction_function}
\end{figure}

The compaction function (first introduced in \cite{Shibata:1999zs}, see also \cite{Harada:2023ffo,Harada:2024trx}) has been found numerically to be useful for characterizing the threshold for black hole formation \cite{Shibata:1999zs,Harada:2015yda,Musco:2018rwt,deltacq,Musco:2020jjb} in a radiation-dominated Universe, in particular using its maximum value.
The characteristics of type $\uI$ and $\uII$ fluctuations are well understood through the compaction function, expressed at leading order in gradient expansion at super-horizon scales as \cite{Harada:2015yda}, 
\begin{equation}
   \mathcal{C}(r) = \frac{2}{3}\left( 1-{\left(1+r \zeta^{\prime}(r)\right)}^2\right) \label{eq:compaction}
 \end{equation}
 in the comoving gauge\footnote{There are two conventions having factor 2 difference from each other. Our definition is twice the original definition.}. 
Fig.~\ref{fig:compaction_function} shows the relation between the compaction function $\mathcal{C}(r)$ and the coordinate radius $r$ for different values of $\beta$ with varying $\mu$. 
Fluctuations of type I are characterized by the existence of one extremal, which corresponds to a maximum at the scale $r_m$
fulfilling the equation $\zeta'(r_m)+r_m \zeta''(r_m)=0$ with the areal radius being a monotonic increasing function of $r$. Fluctuations of type II are characterized by non-monotonic behavior of the areal radius, and there are three extremal points $\mathcal{C}'(r)=0$ given by 
$r_{m,1}$, $r_{m,2}$ and $r_{m,3}$ (see appendix C of \cite{Uehara:2024yyp}), two of which at $r = r_{m1}$ and $r_{m,3}$ correspond to two peaks\footnote{Notice that we can have two peaks in the compaction function with fluctuations of type I with $\mathcal{C}_{\rm peak}<2/3$, see for instance \cite{Escriva:2023qnq}.} 
with the value $2/3$ satisfying $1+r_{m,1,3} \zeta'(r_{m,1,3})=0$, and the rest at $r = r_{m,2}$ to its minimum below $2/3$ fulfilling $\zeta'(r_{m,2})+r_{m,2} \zeta''(r_{m,2})=0$. 
See Fig.~\ref{fig:compaction_function} for the transition from type I to type II as increasing the amplitude. 
Through the transition, the first local maximum $r_m$ is split into two maxima $r_{m,1}$ and $r_{m,3}$.

\begin{figure}[t]
    \centering
    \includegraphics[width=0.6\linewidth]{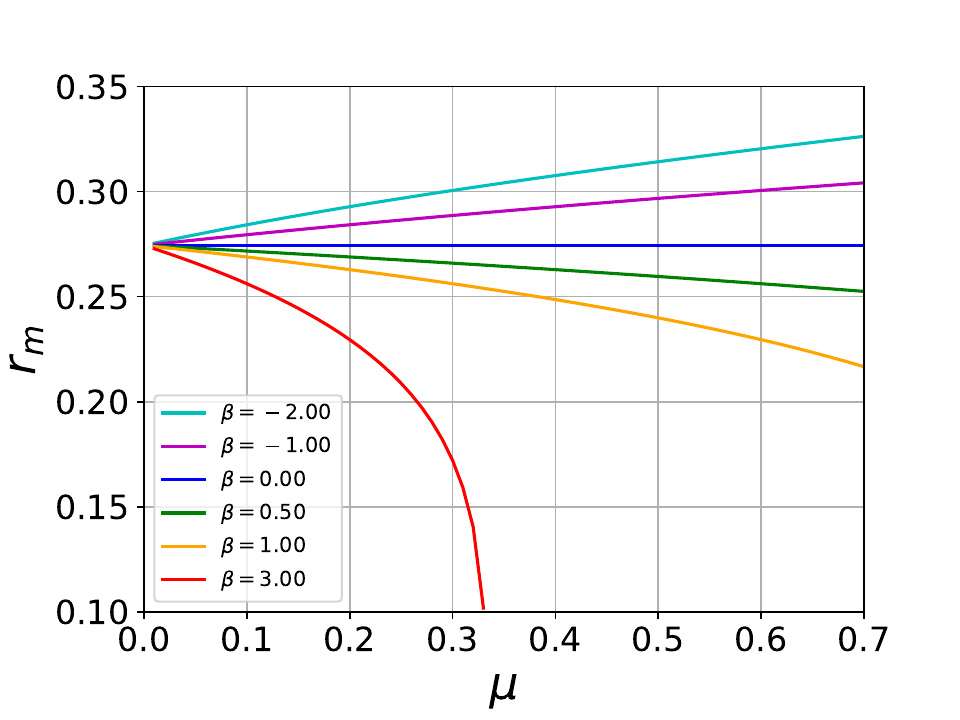}
    \caption{Values of the location of the peak value of the compaction function $\mathcal{C}$ in terms of the amplitude of the fluctuation $\mu<\mu_{II}$.}
    \label{fig:rmmu}
\end{figure}

In Ref.~\cite{deltacq} were proposed analytical estimates to accurately determine the black hole formation threshold in a radiation-dominated universe, accounting for the profile dependence of type I fluctuations. For those cases, the threshold is characterized by the peak value of the compaction function, $\mathcal{C}(r_m)$. In particular, it was found that the average critical compaction function integrated up to the peak $\mathcal{C}(r_m)$ is roughly a universal estimator with value $\bar{\mathcal{C}_c}=2/5$ in a radiation-dominated Universe.  
    More explicitly, $\bar{\mathcal{C}}_c$ taking into account non-Gaussian contribution is defined as \cite{Atal_2020} 
    \begin{equation}
        \bar{\mathcal{C}}_c = \frac{3}{r_{m}^3(\mu,\beta) e^{3\zeta(r_{m}(\mu,\beta))}} \int^{r_{m}(\mu,\beta)}_0 C_{c}(r)\left(1+r\zeta^{\prime}(r)\right)e^{3\zeta(r)}r^2dr, 
        \label{eq:averaged}
    \end{equation}
    where we account for the dependence of $r_m$ on 
    $\mu$ (see Fig.~\ref{fig:rmmu}).

In addition, in \cite{deltacq}, it was discovered that the threshold for black hole formation mainly depends on the shape around the peak of the compaction function (namely $r_m$) through a dimensionless parameter $q=-r_{\rm co,m}^2 \mathcal{C}^{\prime \prime}(r_{\rm co,m})/(4 \mathcal{C}(r_{\rm co,m}))$ that accounts for the curvature of the compaction function profile with $r_{\rm co,m}$ being the comoving coordinate radius.  
In terms of the coordinate system chosen in this work, it is written as
\begin{equation}
\label{eq:parametter}
    q = -\frac{1}{4}r_{m}^2 \frac{\mathcal{C}^{\prime \prime}(r_{m})}{\mathcal{C}(r_m)\left(1-\frac{3}{2} \mathcal{C}(r_{m})\right)}.
\end{equation}
With the help of a specific model profile, one can relate the average compaction $\bar {\mathcal C}$ and $q$ parameter. 
Then, by inverting the averaged formula, 
one can get the following analytical expression of the threshold value for the compaction function as a function of $q$:
\cite{deltacq}
\begin{equation}
    \delta_c(q) = \frac{4}{15}e^{-\frac{1}{q}}\frac{q^{1-\frac{5}{2q}}}{\Gamma\left(\frac{5}{2q}\right) - \Gamma\left(\frac{5}{2q},\frac{1}{q}\right)},
    \label{eq:delta_c_formula}
\end{equation} 
where $\Gamma(x)$ is the Gamma function and $\Gamma(x,y)$ the incomplete Gamma function.
The value of $\mu$ that satisfies the condition $\delta_c(q(\mu)) = \mathcal{C}(\mu,r_m(\beta))$ 
gives the threshold for $\mu$ for a given value of the non-gaussianity parameter $\beta$.
We note that
Eqs.~\eqref{eq:delta_c_formula} and \eqref{eq:averaged} represent the analytical approaches for estimating the threshold for black hole formation under the assumption of fluctuations of type I.

The analytical approach for threshold estimation was found to accurately estimate the thresholds for $\beta \geq 0$ \cite{Atal_2020} when compared with the results from simulations. 
However, when considering the quadratic local template of non-Gaussianities, 
in \cite{Escriva:2022pnz}, it was observed that, for sufficiently large negative values of non-Gaussianities, 
the approach using the $q$ parameter gave rise to much more accurate results than using the averaged $\mathcal{C}$, while both approaches are very accurate for positive non-Gaussianities.
This may be due to the existence of a negative region in the compaction function for sufficiently negative non-Gaussianities.

\section{Numerical evolution}
\label{Num}

In this section, we present 
the initial setting and 
the results of the numerical simulations.

\subsection{Initial data}
\label{numerical set}
As mentioned earlier, the initial conditions for curvature fluctuations
~\eqref{eq:zeta_log} are set on the super-horizon scale. 
At the initial time $t=t_{\rm i}$, we set $a(t_{\rm i})=1$ and $k/H_b(t_{\rm i})=0.1$. The numerical simulations are performed in the domain given by $r<L$ with $k=10/L$ and $H_b(t_{\rm i})=100/L$. 
Since the expression \eqref{eq:zeta_log} is indeterminate at the center, we use the form of the Taylor expansion in the vicinity near the center. 
Additionally, a window function $W(r)$ will be introduced for regions where $r$ is large, which imposes the condition $\zeta(r) = 0$ at the outer boundary $L$ \cite{Windowfunction} 
necessary to fulfill the boundary conditions required in our specific numerical code. 
In appendix \ref{apendix_window}, we briefly discuss the effects of the window function in our numerical results.
Consequently, the specific expression for the initial curvature fluctuation $\zeta$ is given as follows: 
\begin{align}
    \zeta =
        -\frac{1}{\beta}\ln(1-\beta\mu\sinc(kr))W(r),
        \label{eq:zeta_W}
\end{align}
where the explicit form of the window function is given by 
\begin{align}
    W(r) = 
    \begin{cases}
        1 & \text{for $0\leq r\leq r_{\text{inner}}$}\\
        1 - \frac{((r_{\text{inner}}-r_{\text{outer}})^6 - (r_{\text{outer}} - r)^6)^6}{(r_{\text{inner}} - r_{\text{outer}})^{36}} & \text{for $r_{\text{inner}} \leq r \leq r_{\text{outer}}$}\\
        0& \text{for $r_{\text{outer}} \leq r$}
    \end{cases}
\end{align}
with $r_{\rm inner}/L=0.45$ and $r_{\rm outer}/L = 0.77$. 

\begin{figure}[b]
\centering
\begin{minipage}[t]{0.45\linewidth}
    \centering
    \includegraphics[width=\linewidth]{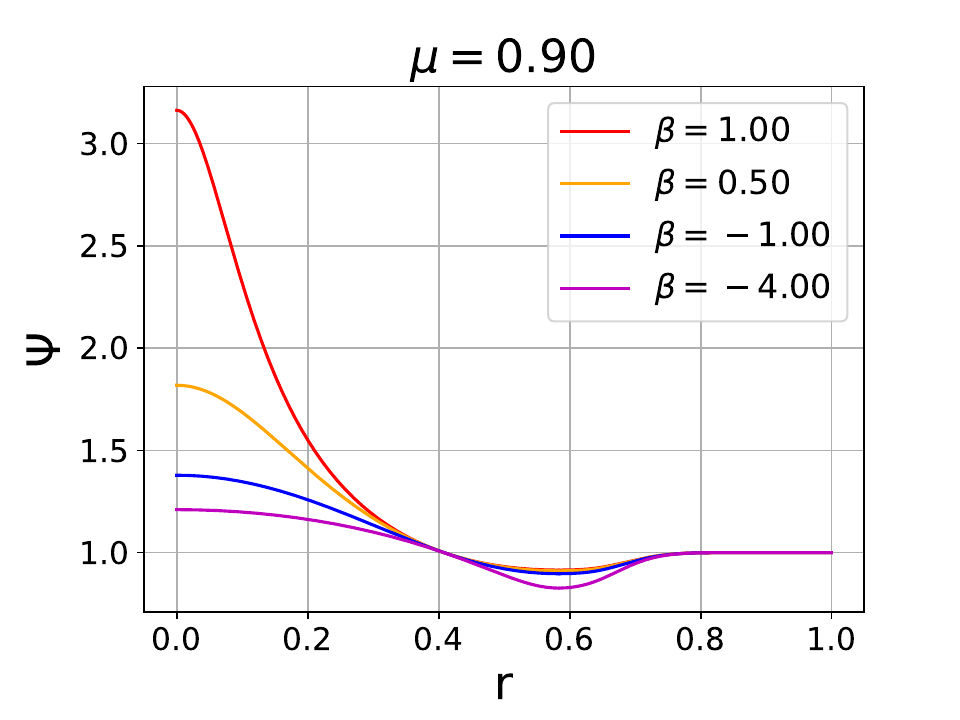}    
\end{minipage}
\begin{minipage}[t]{0.45\linewidth}
    \centering
    \includegraphics[width=\linewidth]{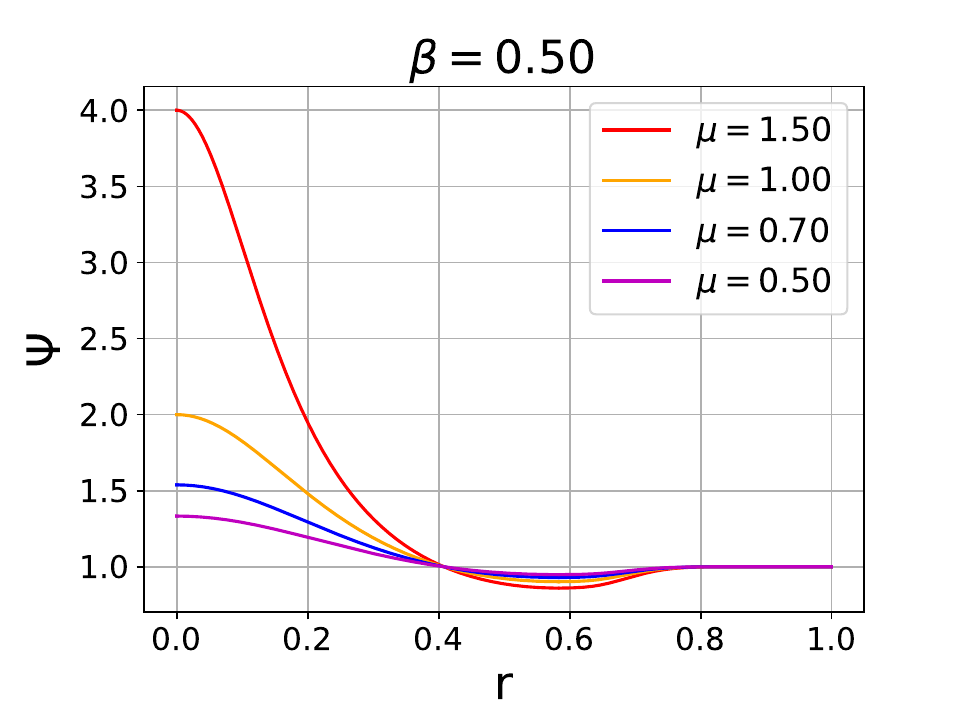} 
\end{minipage}
    \caption{
    The initial profile of $\Psi$ for $\mu = 0.90$ with $\beta = -4.00$, $-1.00$, $0.50$ and $1.00$ (left), and for $\beta = 0.50$ with $\beta = 0.50$, $0.70$, $1.00$ and $1.50$ (right).
    }\label{fig:psi}
\end{figure}

We show the initial profile of conformal factor $\Psi = e^{\zeta/2}$ 
for $\mu=0.9$ with varying $\beta$ (left panel), and for $\beta=0.5$
with varying $\mu$ (right panel)
in Fig.~\ref{fig:psi}.
From Fig.~\ref{fig:psi}, it
can be observed that the amplitude of $\Psi$ increases as $\mu$ or $\beta$ increases.

\subsection{Time evolution of the fluctuations and its collapse/dispersion}
\begin{figure}[htbp]
    \centering
    \begin{minipage}[ht]{0.32\columnwidth}
        \includegraphics[width = \columnwidth]{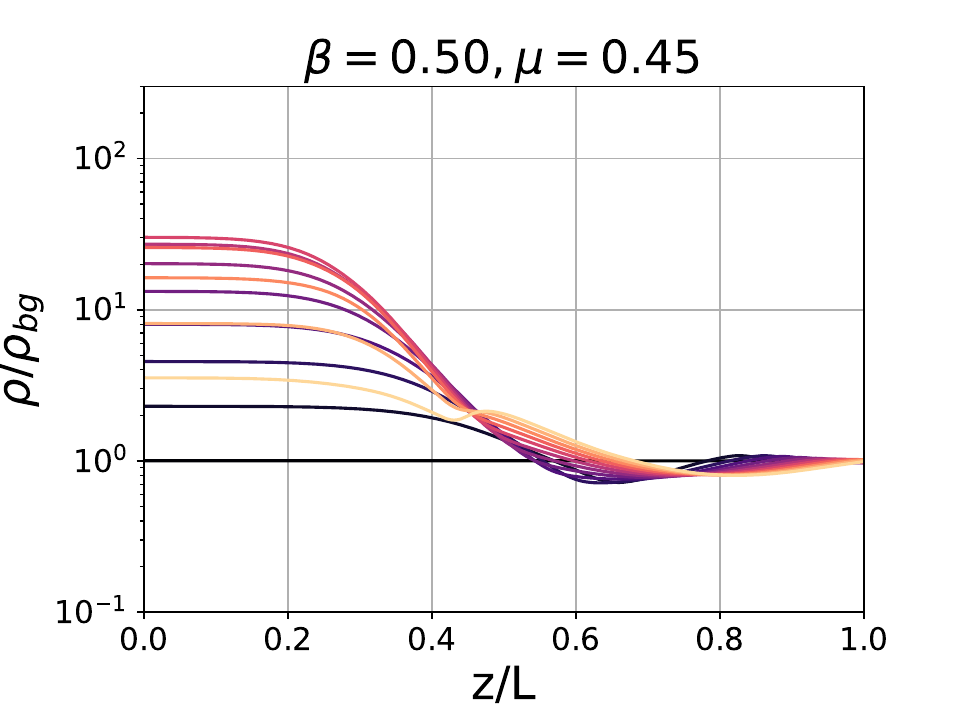}
    \end{minipage}
    \begin{minipage}[ht]{0.32\columnwidth}
        \includegraphics[width = \linewidth]{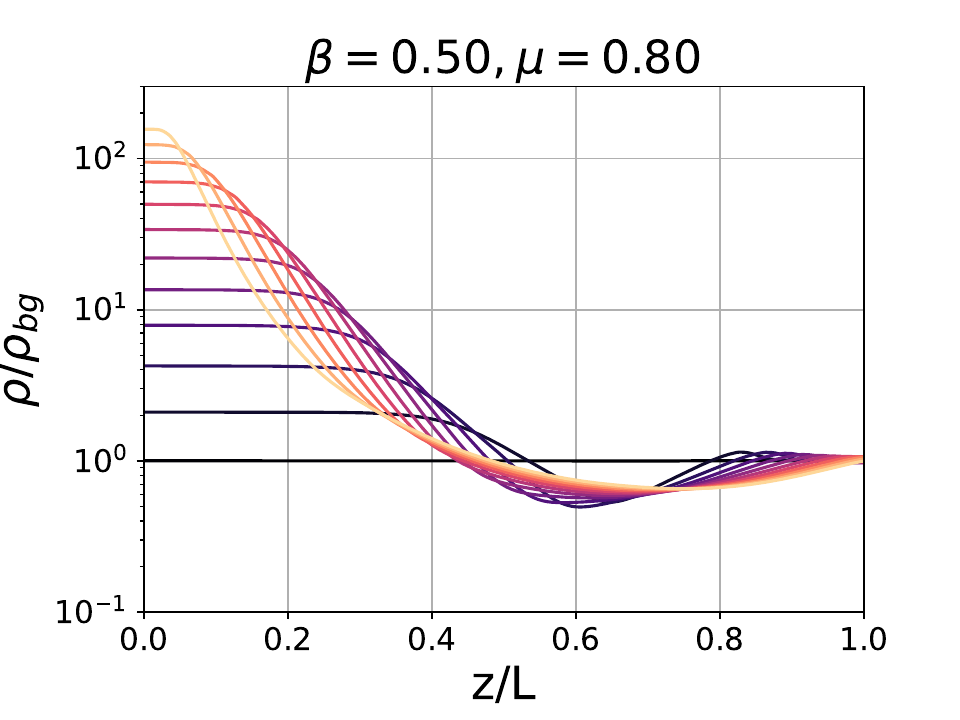} 
    \end{minipage}
    \begin{minipage}[ht]{0.32\columnwidth}
        \includegraphics[width =\linewidth]{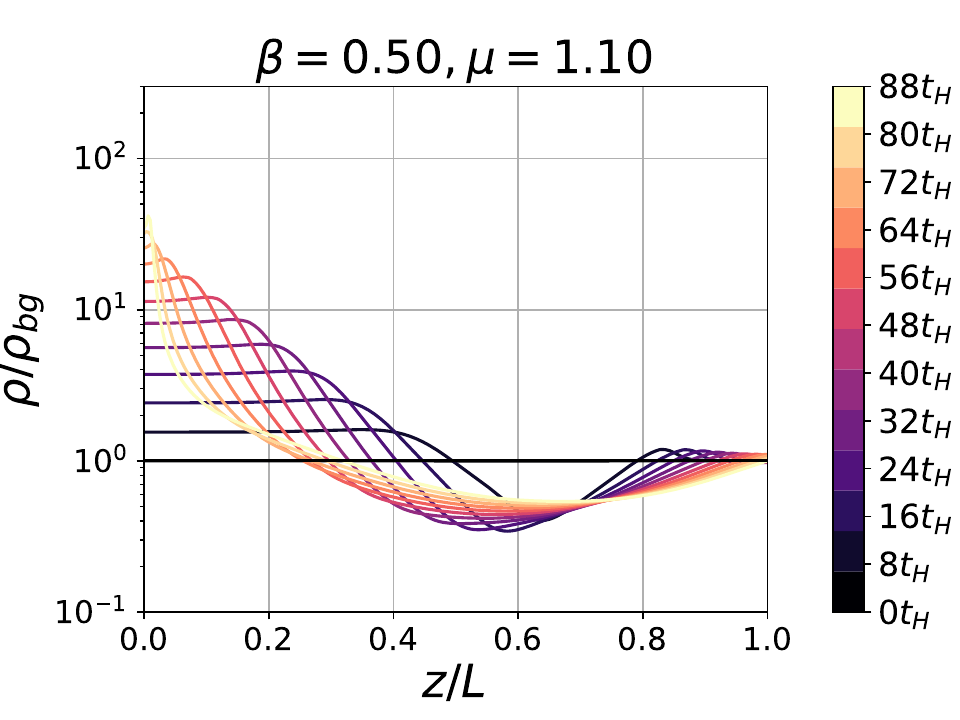}
    \end{minipage}
    \centering
    \begin{minipage}[ht]{0.32\columnwidth}
        \includegraphics[width = \columnwidth]{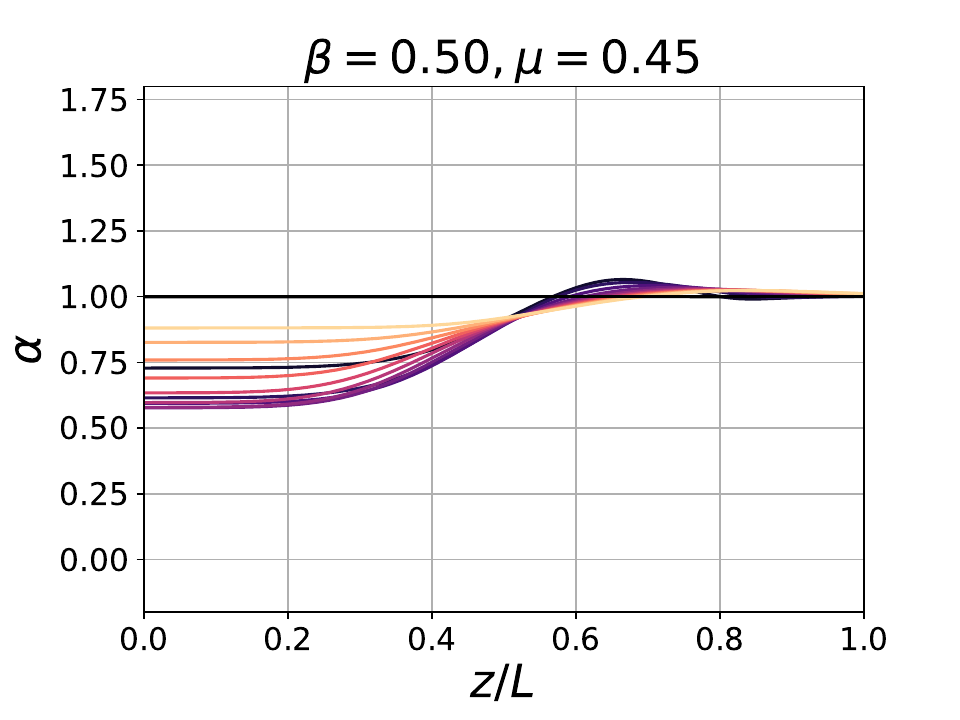}
    \end{minipage}
    \begin{minipage}[ht]{0.32\columnwidth}
        \includegraphics[width = \linewidth]{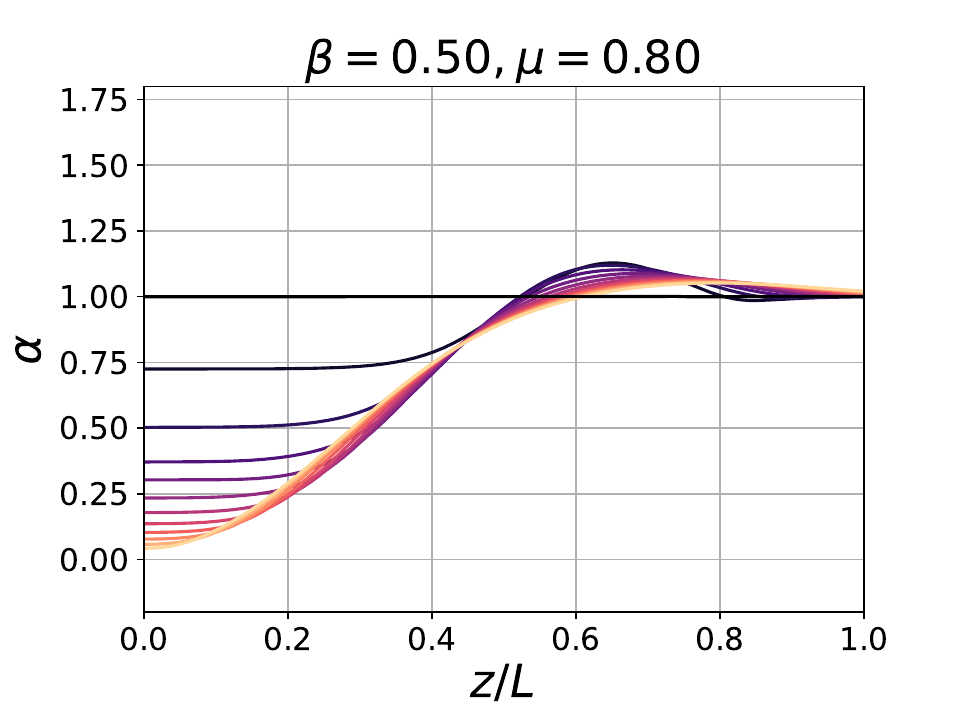} 
    \end{minipage}
    \begin{minipage}[ht]{0.32\columnwidth}
        \includegraphics[width =\linewidth]{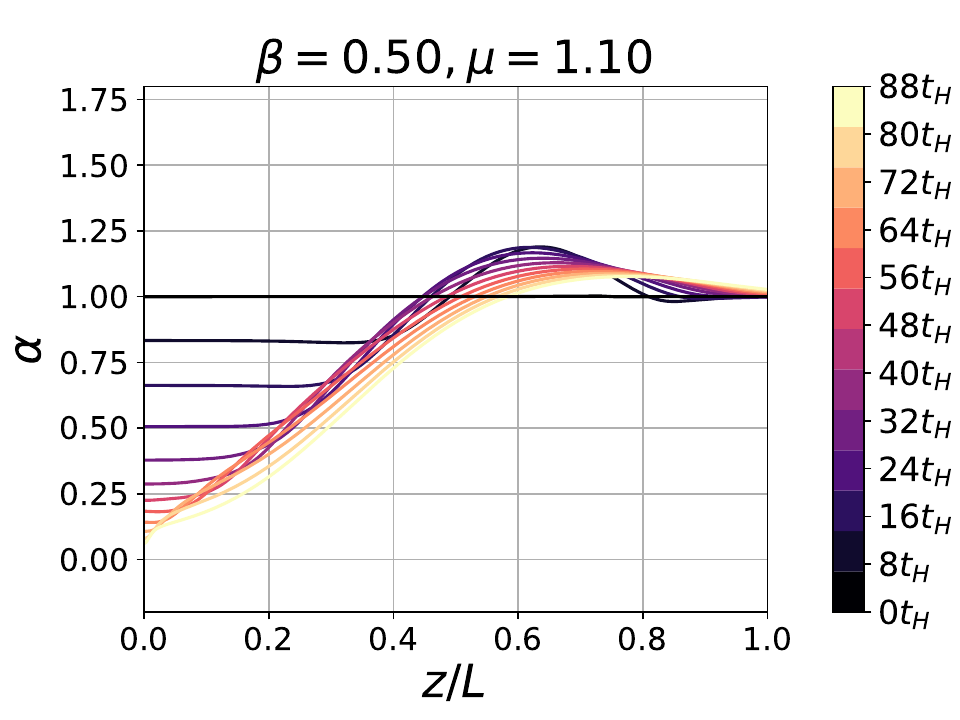}
    \end{minipage}
    \caption{
    The time evolution of the fluid density contrast $\rho/\rho_{bg}$ (upper-panels) and lapse function $\alpha$ (lower-panels) for $\beta = 0.50$ of $\mu = 0.45$ (no PBH formation case), $\mu = 0.80$, and $1.10$ (PBH formation case).}\label{fig:lapsedelta_betafix}
\end{figure}

Depending on the fluctuation amplitude
$\mu$, the initial fluctuations form black holes or disperse on the FLRW background.
To calculate the time evolution generated from the initial functional form of $\zeta$ ~\eqref{eq:zeta_W} during the radiation-dominated era with the equation of state parameter $w = 1/3$, we will use the numerical code COSMOS-S \cite{COSMOS}. 
In the calculations, instead of using the coordinate radius $r$, we will use the scale-up coordinate $z$, which is related to $r$ as 
\begin{equation}
    r = z- \frac{\eta}{1+\eta}\frac{L}{\pi}\sin\left(\frac{\pi}{L}z
    \right)
\end{equation} 
with $\eta = 20$.
We show the time evolution of the contrast of the fluid energy density $\rho/\rho_{bg}$ and lapse function $\alpha$ for three values of $\mu$ ($0.45$, $0.80$ and $1.10$) with $\beta=0.5$ in Fig.~\ref{fig:lapsedelta_betafix}. 

For the case of $\mu =0.45$ (left), the density contrast at the center initially increases but returns to a smaller value after $t\sim40t_{H}$.
Here, $t_{H}$ is the horizon reentry time defined as $k^{-1} = (aH_{b})^{-1}|_{t = t_{H}}$. The value of the lapse at the center first decreases until it reaches a minimum value, and increases after that. 
On the other hand, for $\mu = 0.8$ (middle) and $\mu = 1.1$ (right) cases, which respectively correspond to type I and II fluctuations, the density at the center continues to increase and forms a peak, whereas, for the lapse function, it continues to decrease to $\alpha\sim 0$. 
We do not
appreciate significant qualitative differences for the different values of $\beta$ considered.
These behaviors are quite common when we simulate gravitational collapse (see, e.g., \cite{Musco:2012au,Escriva:2019nsa,Yoo:2021fxs,Uehara:2024yyp} for PBH formation),
and useful for roughly checking if a fluctuation collapses forming an apparent horizon or not.

\subsection{Type A/B primordial black hole and threshold values}
\begin{figure}[htbp]
    \centering
    \begin{minipage}[h]{0.32\columnwidth}
        \includegraphics[width = \columnwidth]{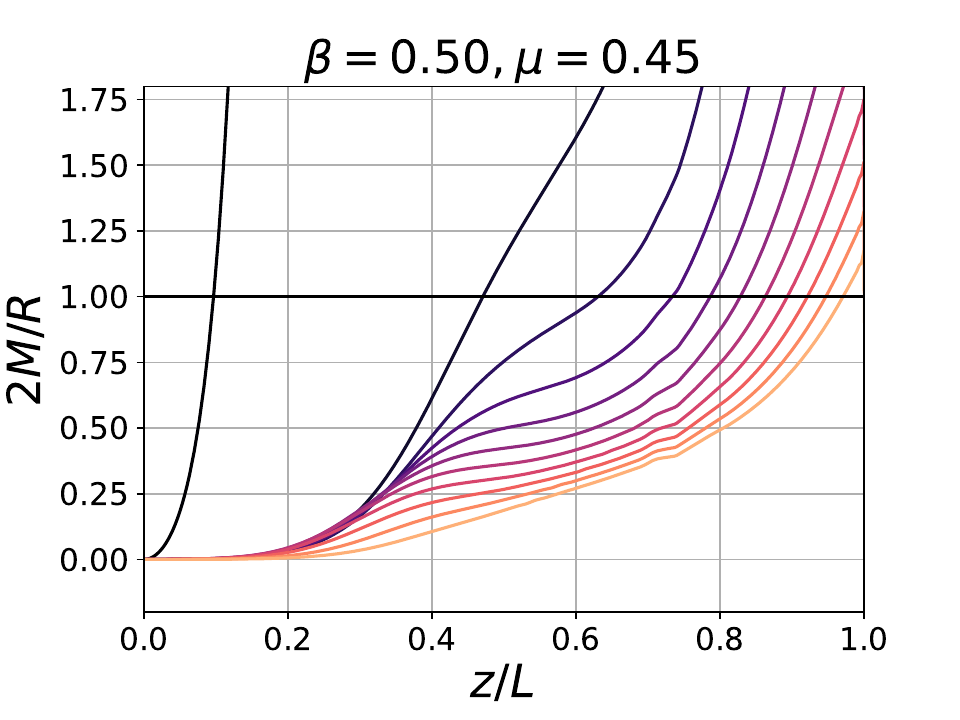}
    \end{minipage}
    \begin{minipage}[h]{0.32\columnwidth}
        \includegraphics[width = \linewidth]{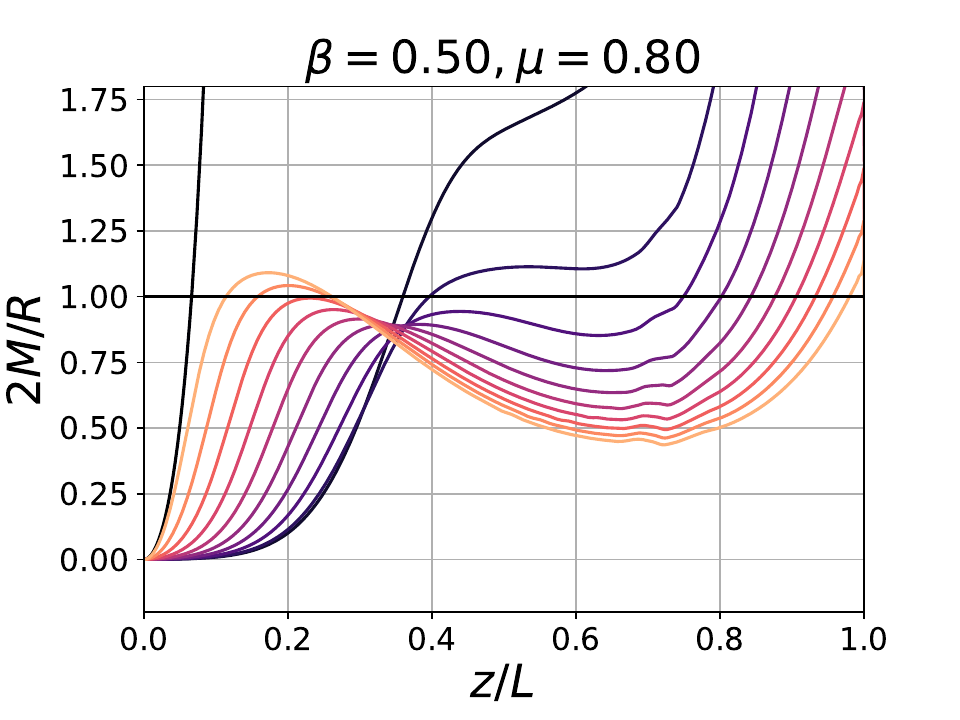} 
    \end{minipage}
    \begin{minipage}[h]{0.32\columnwidth}
        \includegraphics[width =\linewidth]{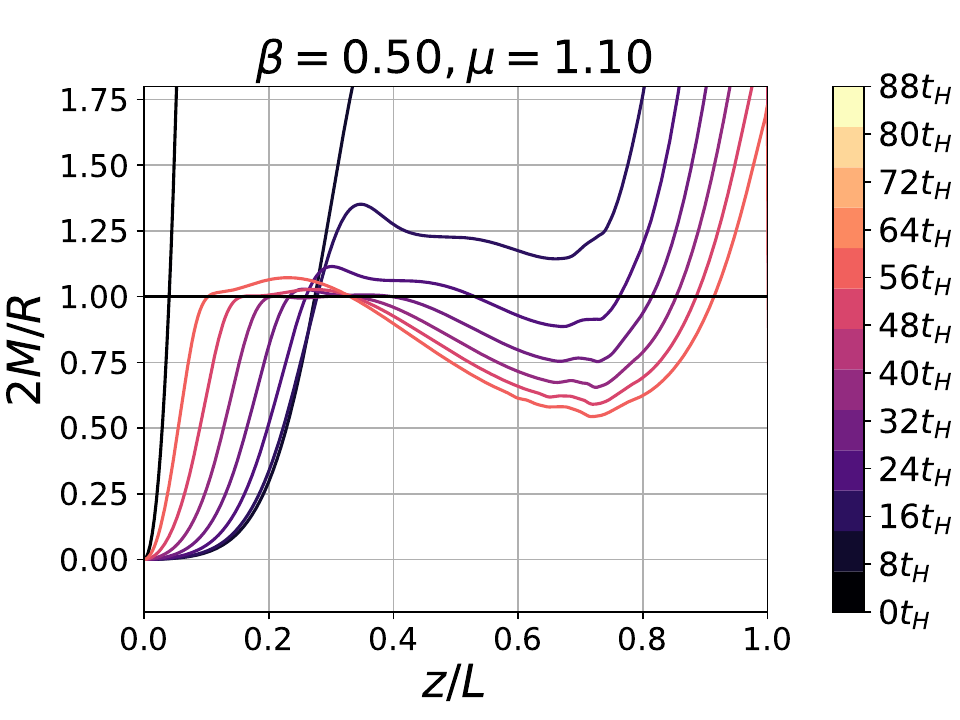}
    \end{minipage}
    \centering
    \begin{minipage}[h]{0.32\columnwidth}
        \includegraphics[width = \columnwidth]{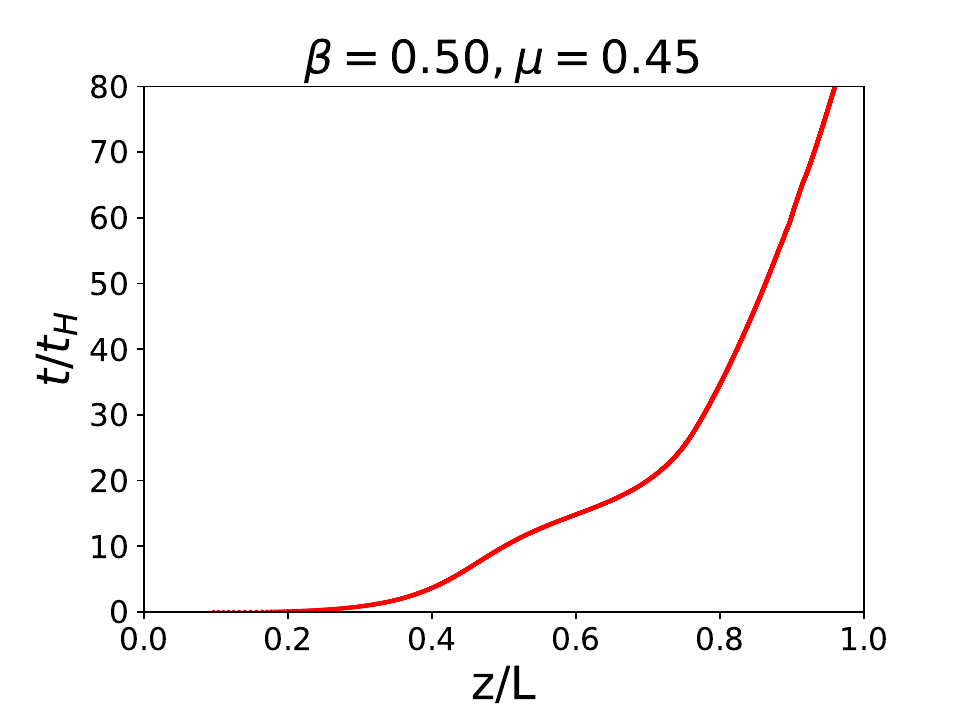}
    \end{minipage}
    \begin{minipage}[h]{0.32\columnwidth}
        \includegraphics[width = \linewidth]{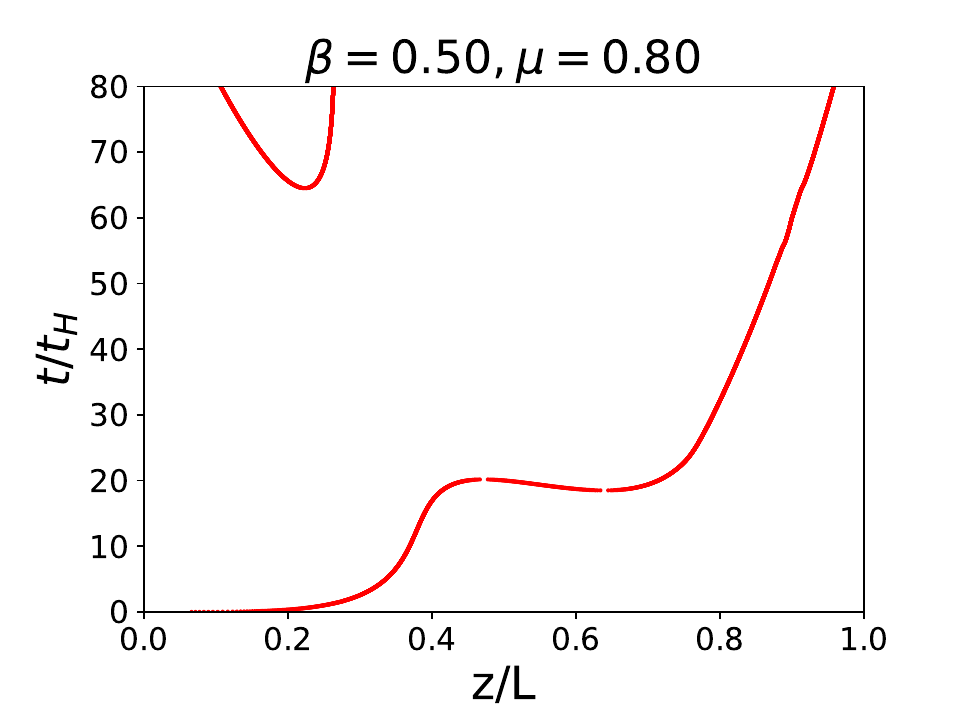} 
    \end{minipage}
    \begin{minipage}[h]{0.32\columnwidth}
        \includegraphics[width =\linewidth]{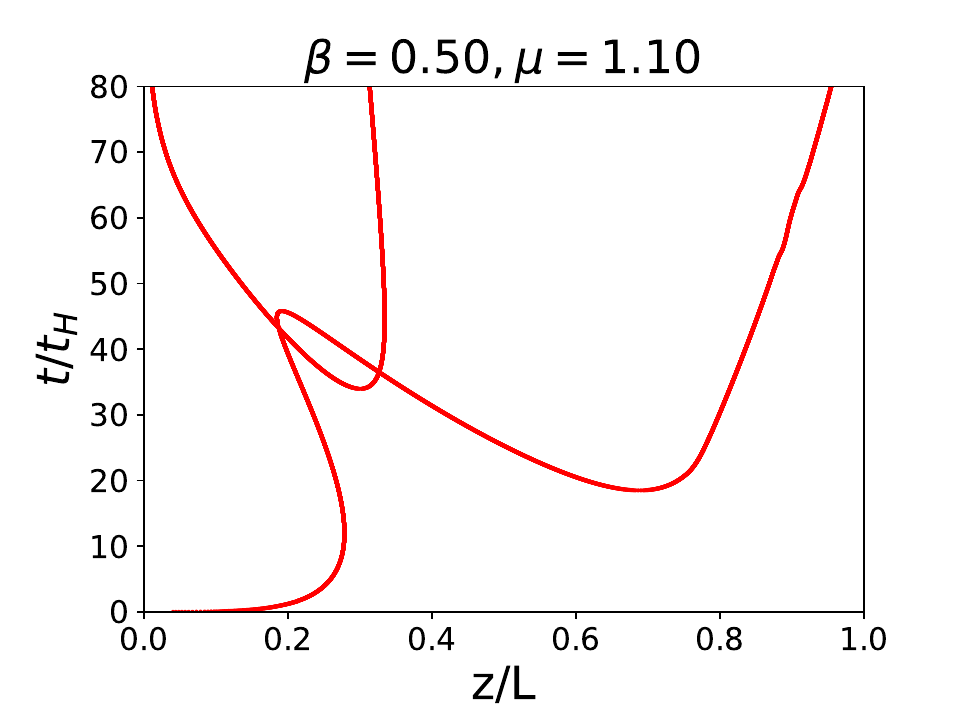}
    \end{minipage}
    \caption{
    Snapshots of the $2M/R$ (upper-panels) and the trajectories of the trapping horizons in time (lower-panels) for a fixed value of $\beta = 0.50$ and changed $\mu = 0.45, 0.80$ and $1.10$. $t_H$ refers to the time of horizon reentry.}
    \label{fig:PBHtype_betafix}
\end{figure}

\begin{figure}[htbp]
    \centering
    \begin{minipage}[t]{0.32\columnwidth}
        \includegraphics[width = \columnwidth]{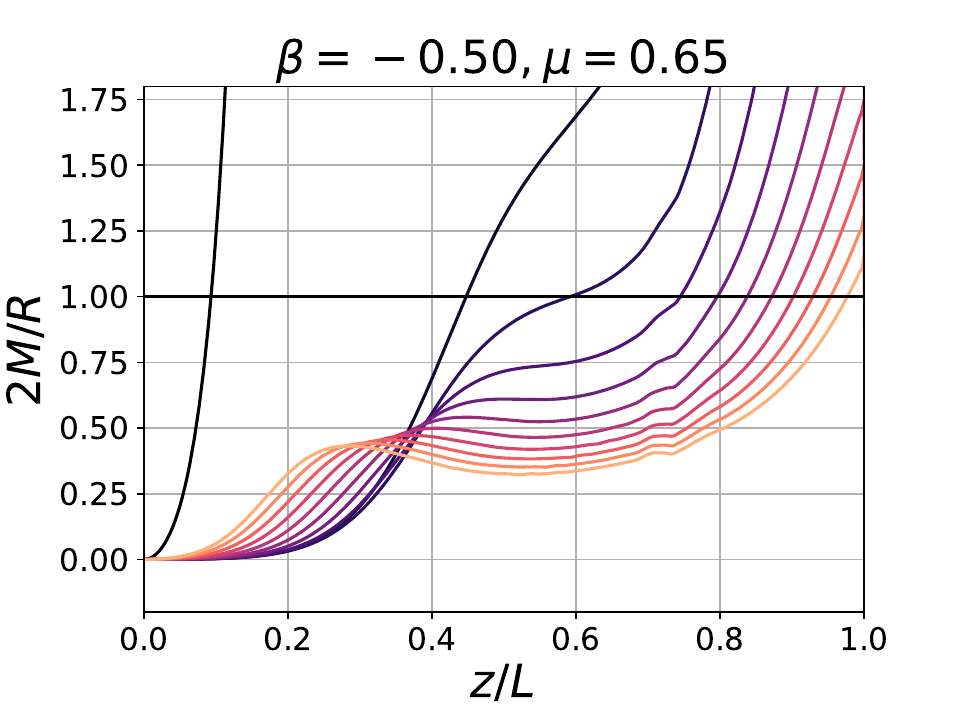}
    \end{minipage}
    \begin{minipage}[t]{0.32\columnwidth}
        \includegraphics[width = \linewidth]{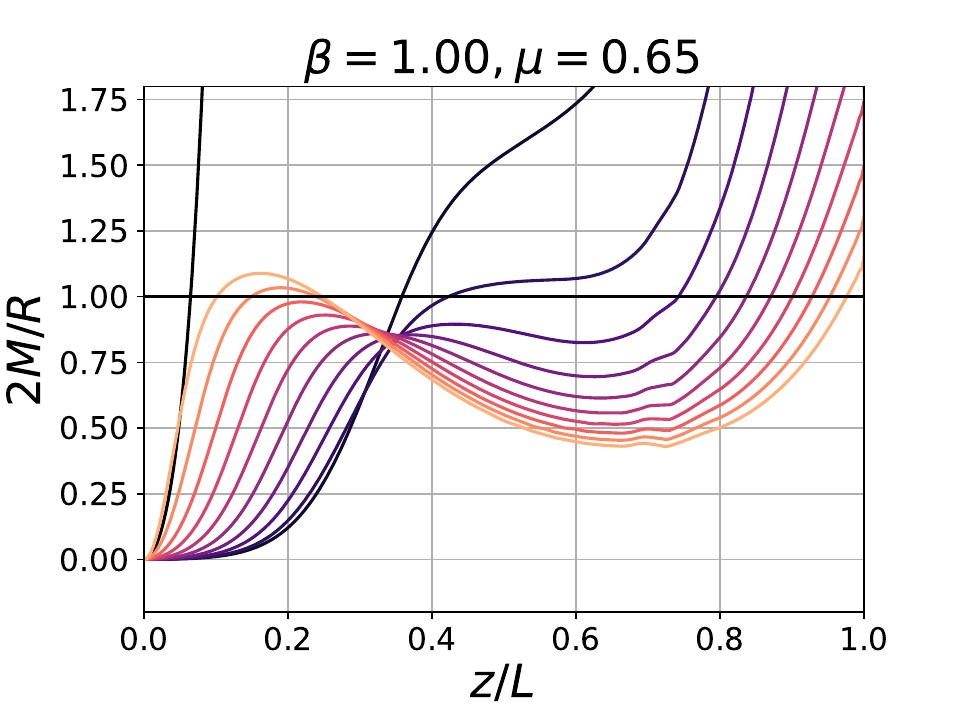} 
    \end{minipage}
    \begin{minipage}[t]{0.32\columnwidth}
        \includegraphics[width =\linewidth]{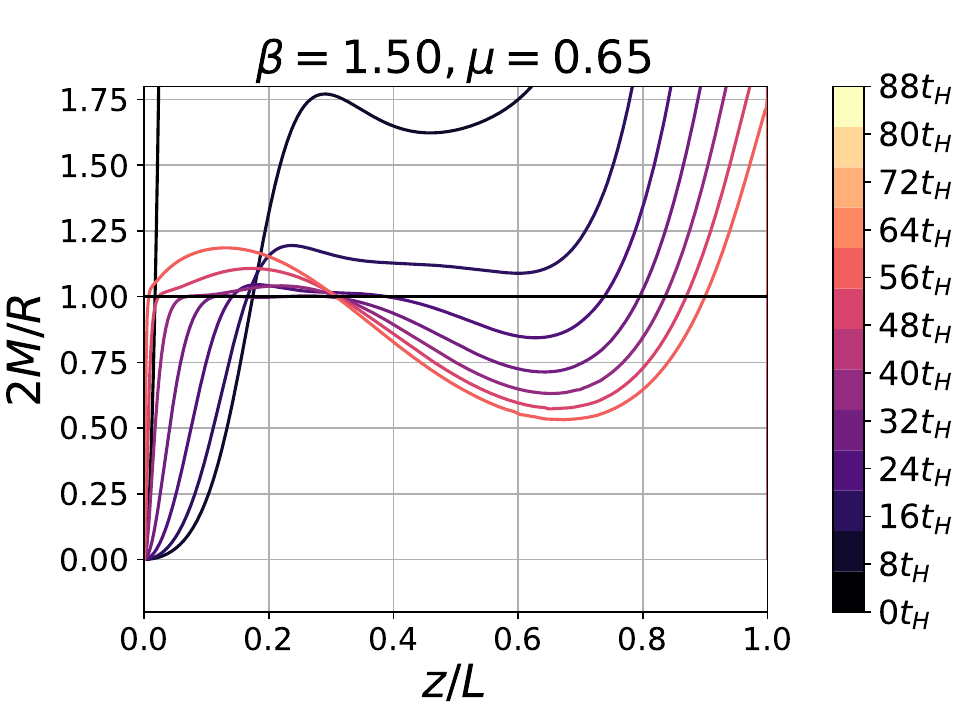}
    \end{minipage}
    \centering
    \begin{minipage}[t]{0.32\columnwidth}
        \includegraphics[width = \columnwidth]{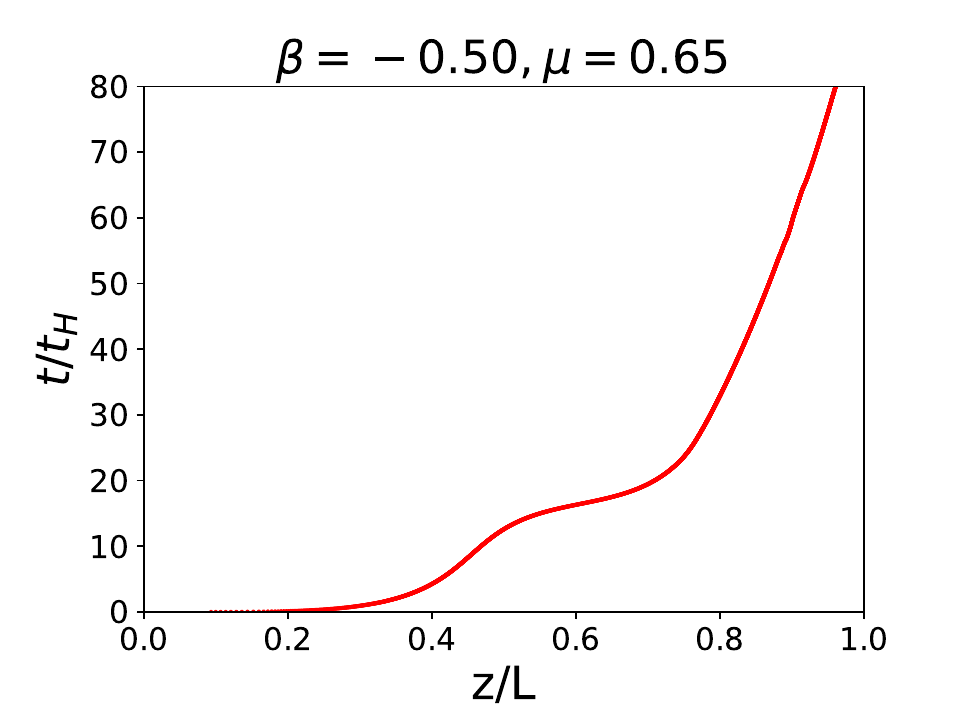}
    \end{minipage}
    \begin{minipage}[t]{0.32\columnwidth}
        \includegraphics[width = \linewidth]{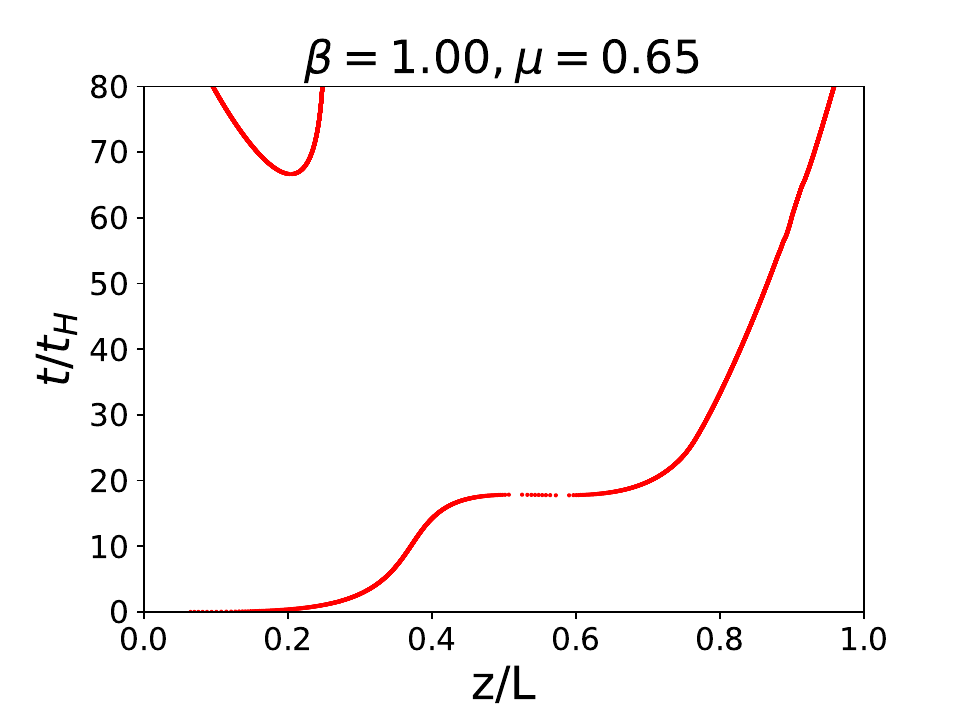} 
    \end{minipage}
    \begin{minipage}[t]{0.32\columnwidth}
        \includegraphics[width =\linewidth]{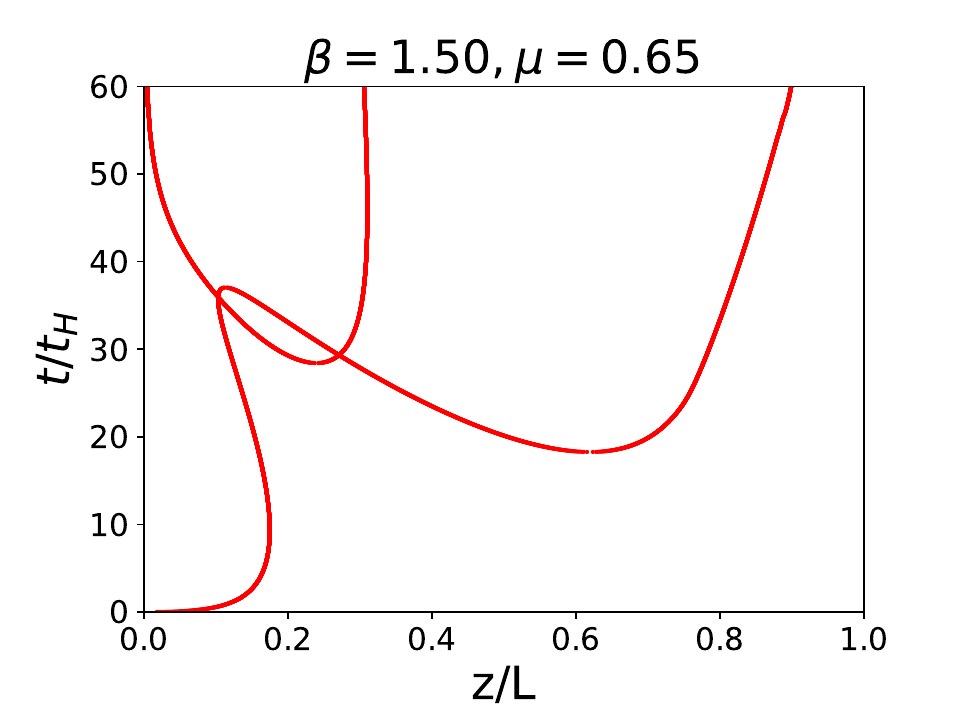}
    \end{minipage}
    \caption{Snapshots of the $2M/R$ (upper-panels) and the trajectories of the trapping horizons in time (lower-panels) for a fixed value of $\mu = 0.65$ and changed $\beta = -0.50, 1.00$ and $1.50$.}\label{fig:PBHtype_mufix}
\end{figure}

We follow the classification of PBHs introduced in \cite{Uehara:2024yyp}, for which we refer the readers for more details.
The crucial difference between type A and type B is the existence of the bifurcating trapping horizon 
which is described by the intersection points of the two trapping horizon trajectories in Fig.~\ref{fig:PBHtype_betafix} and Fig.~\ref{fig:PBHtype_mufix},
where the trapping horizons are defined by the surfaces that satisfy the compactness $2M/R = 1$ with $M$ being the Misner-sharp mass \cite{Yoo:2022mzl,bubble}.
We can confirm that we start to observe type A PBH formation increasing the initial amplitude, and further increasing the amplitude, we observe type B PBH formation.

\begin{figure}[t]
    \centering
    \includegraphics[width = 0.8\columnwidth]{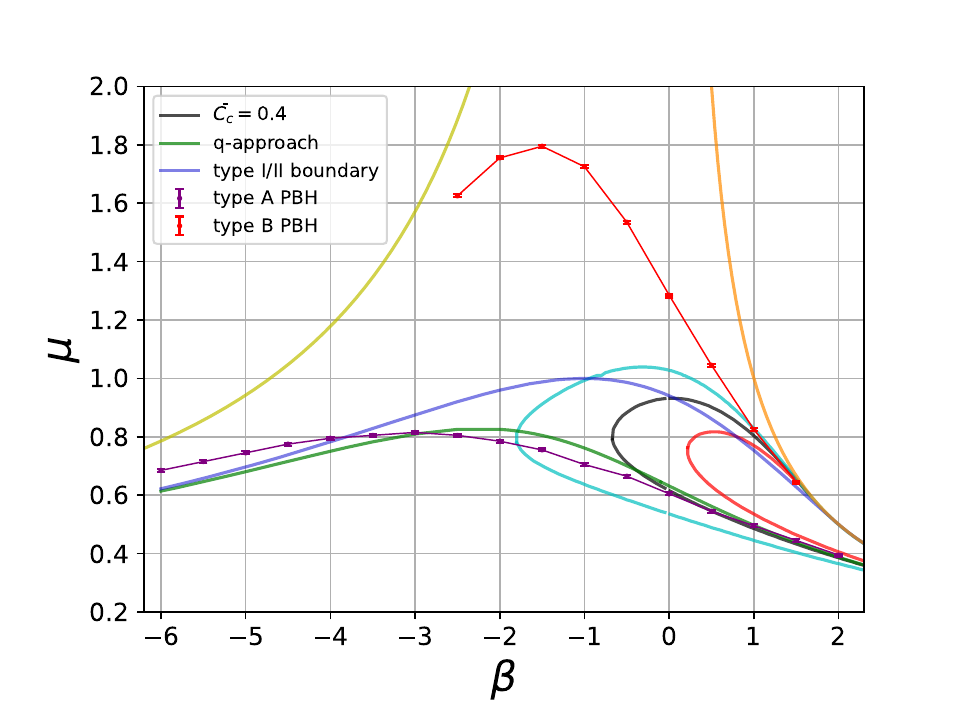}
    \caption{PBH phase diagram
    in terms of $\mu(\beta)$. The blue line corresponds to the boundary between type I and II fluctuations. The orange and yellow lines correspond to $1/\beta$ and $-3\pi/2\beta$, respectively. 
    The purple and red points represent the numerical results of the threshold for type A/B PBH formation. 
    We numerically find no PBH formation with the parameter sets corresponding to the lower edges of the error bars on the purple points, while we find type A PBH formation for the upper edges of them.
    Similarly, for the upper/lower edges of the error bar on the red points, we find type B/A PBH formation. 
     The green solid line is the value inferred from the analytical estimation with the $q$ approach.
    The black line represents the analytical estimation obtained from the averaged critical compaction function approach ($\bar{\mathcal{C}}_c = 0.4$), while the red and cyan lines denote represent values deviating by $\pm 5 \%$ from this estimate, respectively (red line $\bar{\mathcal{C}}_c = 0.42$, cyan line $\bar{\mathcal{C}}_c = 0.38$).
    }
    \label{fig:mucbeta}
\end{figure}

By making several simulations for the different values of $\beta$, 
we can obtain the threshold values for black hole formation (when the fluctuation collapses/disperses) and distinguish between type A/B PBH.
Fig.~\ref{fig:mucbeta} shows 
the summary of the {\em phase diagram} on the $\beta$-$\mu$ plane with our numerical results.
The solid blue line 
shows
the boundary between type I and II fluctuations, namely $\mu_{II}$, derived based on Eq.~\eqref{eq:mu22}
, which exhibits a maximum at $\beta \approx -1$ and decreases as the value of $\beta$ moves away from the maximum point.
The solid orange line denotes $\mu=1/\beta$, the boundary that separates the adiabatic channel of PBH production from the bubble channel~\cite{Atal_2020,bubble} indicated by the divergence of $\zeta$ at $r=0$.
 The yellow line instead denotes the curve $\mu = -3\pi/(2\beta)$, which specifies the limit for what the log-relation Eq.~\eqref{eq:zeta_log} diverges for negative values of $\zeta_G$ 
at the radius $r=3\pi/(2k)$. 
This behavior may be avoided by extending the region of the window function to cover the radius $r=3\pi/(2k)$. However, the late time evolution may be sensitive to the profile at such a relatively large radius. Therefore, for simplicity, we do not discuss the cases of the parameter region above the yellow line further in this paper.

We numerically found the threshold values of the initial amplitude $\mu$ for type A/B PBHs indicated by the purple and red points in Fig.~\ref{fig:mucbeta}, respectively. 
We should notice that for $\beta\lesssim -2.5$, 
we cannot find the threshold for type A/B PBHs within the region we can compute, due to a divergence of the log template Eq.~\eqref{eq:zeta_log} for sufficiently negative values of $\zeta_G$ (see the exclusion region denoted in Fig.~\ref{fig:mucbeta})

\begin{figure}[htbp]
    \centering
    \includegraphics[width = 0.42\columnwidth]{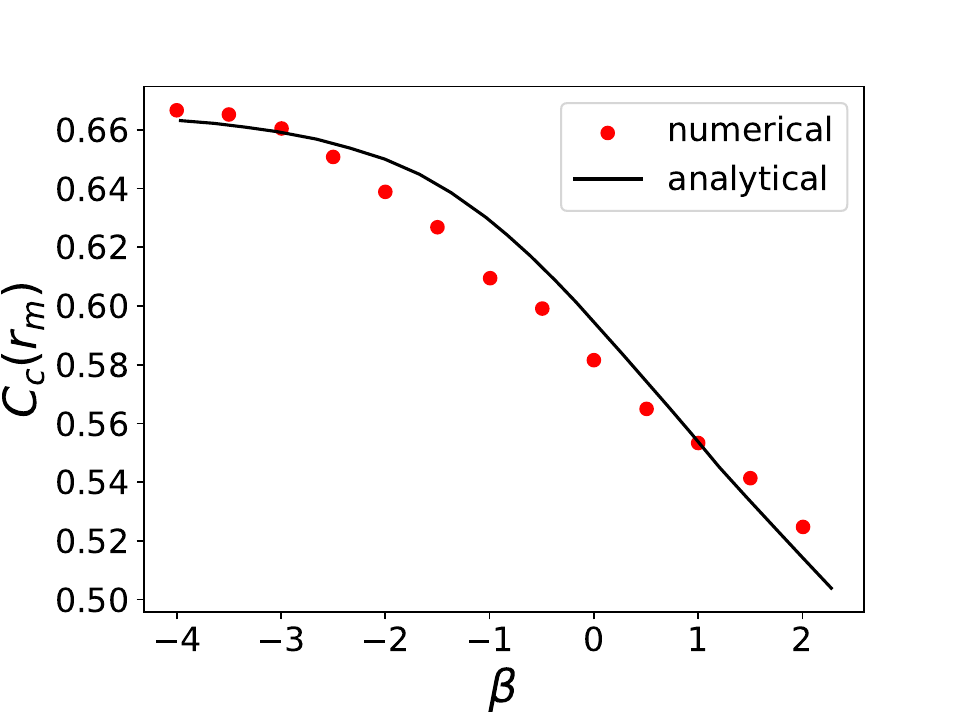}
    \includegraphics[width = 0.4\columnwidth]{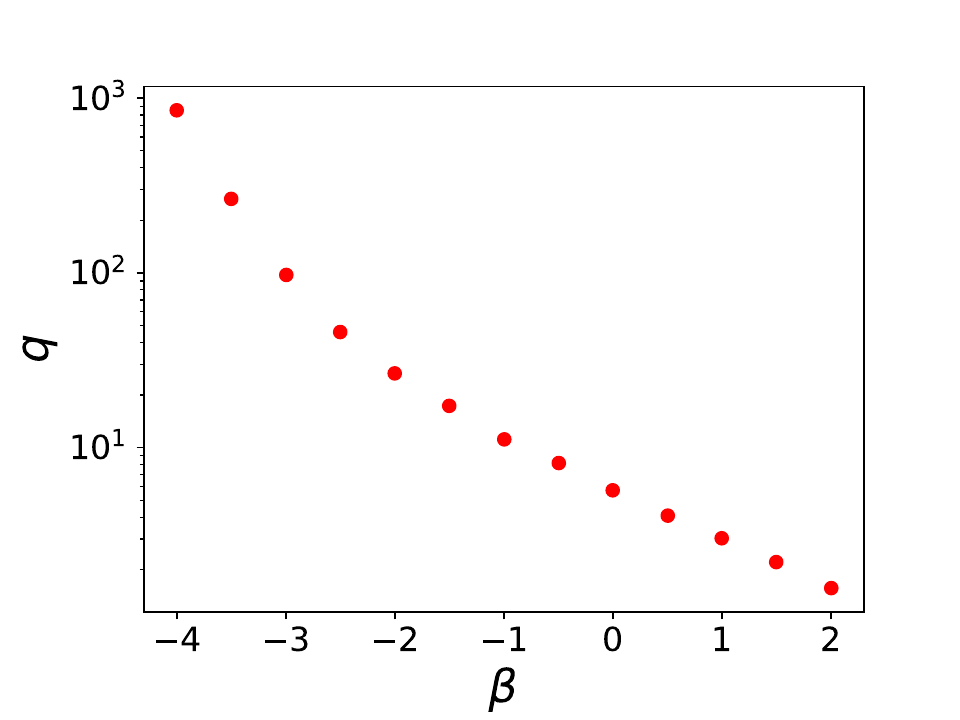}
    \caption{Left-panel: Values of the critical compaction function at its peak value $\mathcal{C}_c(r_m)$ from the numerical results (red points) and analytical ones using Eq.~\eqref{eq:delta_c_formula} (solid line) in terms of $\beta$. Right-panel: Values of the $q$ parameter in terms of $\beta$.} 
    \label{fig:analitical_c}
\end{figure}

For $\beta\geq 0$, we recover the numerical results for the threshold of black hole formation 
obtained in \cite{Atal_2020}\footnote{Notice that in \cite{Atal_2020} a power spectrum with a sharply peaked spectrum was used, rather than monochromatic.}. 
It is interesting to notice that, in this parameter range for $\beta$, when the non-Gaussian parameter $\beta$ is increased, the threshold that distinguishes type A/B PBH becomes smaller. 
This correlation may be related to the role of pressure gradients during the gravitational collapse. More specifically, the correlation 
may originate from the fact that the $q$ parameter decreases with increasing $\beta$ as is shown in Fig.~\ref{fig:analitical_c},
indicating the reduction of pressure gradients. 
We find that for sufficiently large $\beta$, the gap between the threshold of type A/B PBH and that of fluctuations of type II vanishes.

Finally, comparing the analytical estimates with the numerical results,
we see good agreement when using the $q$-approach Eq.~\eqref{eq:parametter} for $\beta  \gtrsim -4$, 
 with deviations within $2-3 \%$. 
 In the case of the averaged $\mathcal{C}$, we find a good agreement for $\beta \geq 0$ as shown in \cite{Atal_2020}, 
 but for negative large $\beta<0$, the deviation is larger. 
 In particular, we account for a deviation of $5-6 \%$ for $\beta \approx -2$\footnote{
 This is much better than the case explored in \cite{Escriva:2022pnz} with the quadratic template due to the existence of a negative region of mass excess. For the log template, we don't find a negative mass excess region at $r < r_m$.}. Interestingly and against expectations, if the value of $\beta$ becomes too small, specifically when $\beta \lesssim -4.0$, the threshold for type A PBH formation lies in the type $\uII$ region (see also~\cite{Inui}). This means that there are fluctuations of type $\uII$ that will not form PBHs, indicating the presence of what can be termed as "type $\uII$ no PBH". For those cases, we have a minimum value of the compaction function at the radius $r_m=r_{m,2}$ together with two neighbour maximums at $r_{m,1}$ and $ r_{m,3}$. The exploration of the criteria for black hole formation and analytical estimates to account for cases of type II fluctuations is left for future research.

\begin{figure}
    \centering
    \includegraphics[width=0.8\linewidth]{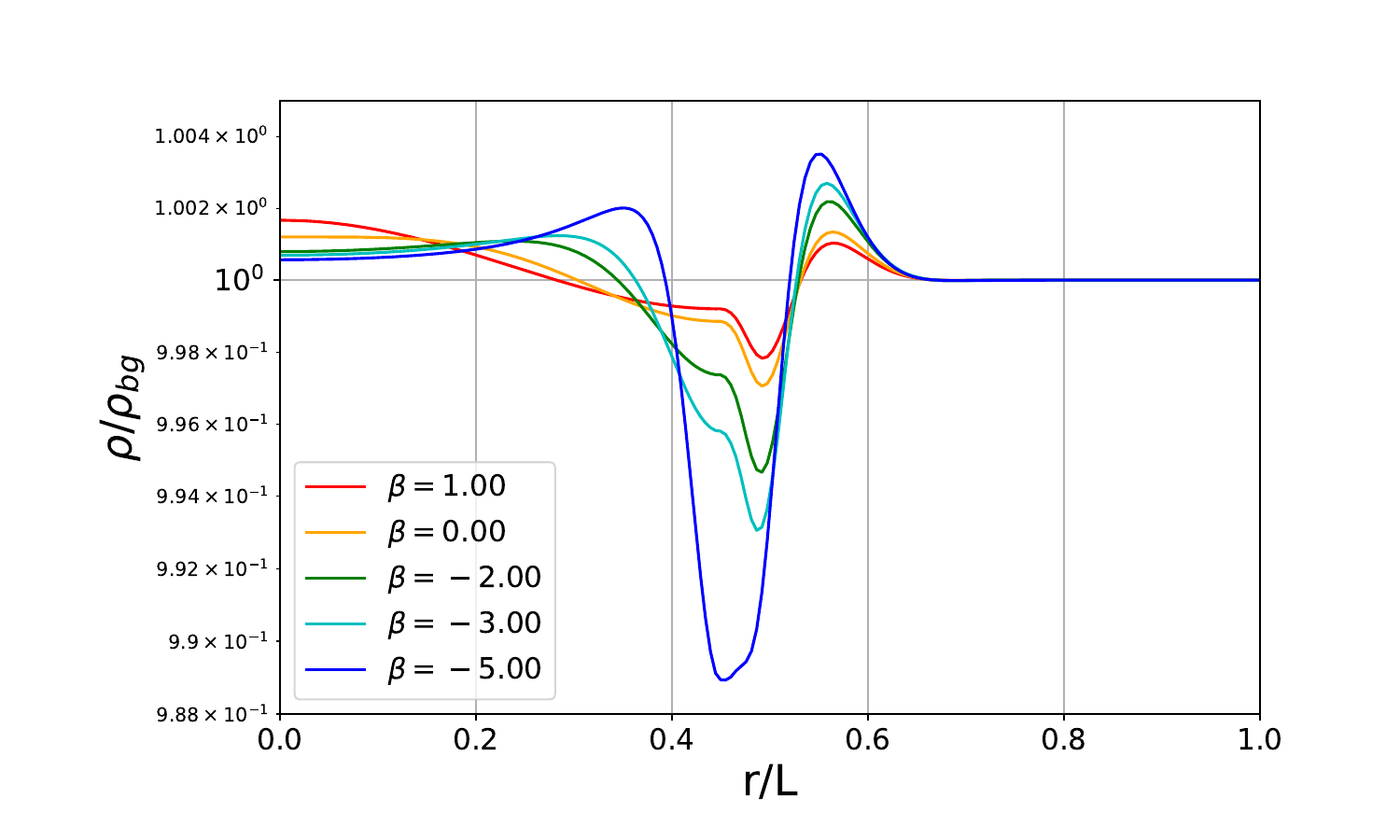}
    \caption{The initial profile of $\rho/\rho_{bg}$ at the critical amplitude for black hole formation
    for $\beta=-5$, $-3$, $-2$, $0$ and $1$.
    }
    \label{fig:initialdelta}
\end{figure}

\begin{figure}[t]
    \centering
    \begin{minipage}[h]{0.40\linewidth}
        \includegraphics[width = \linewidth]{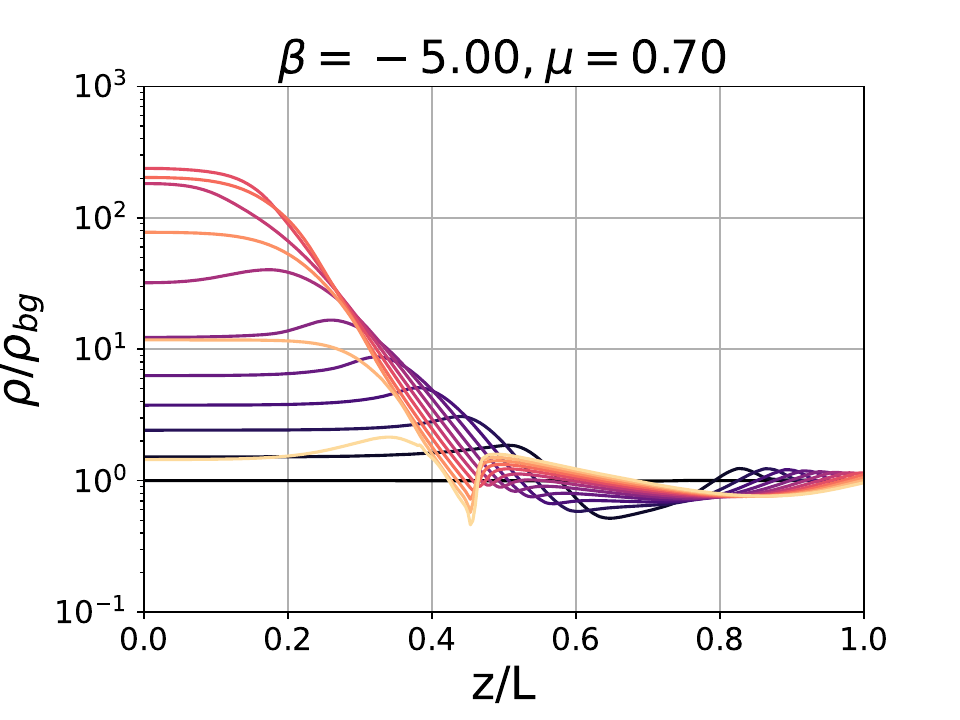}
    \end{minipage}
    \begin{minipage}[h]{0.40\linewidth}
    \includegraphics[width = \linewidth]{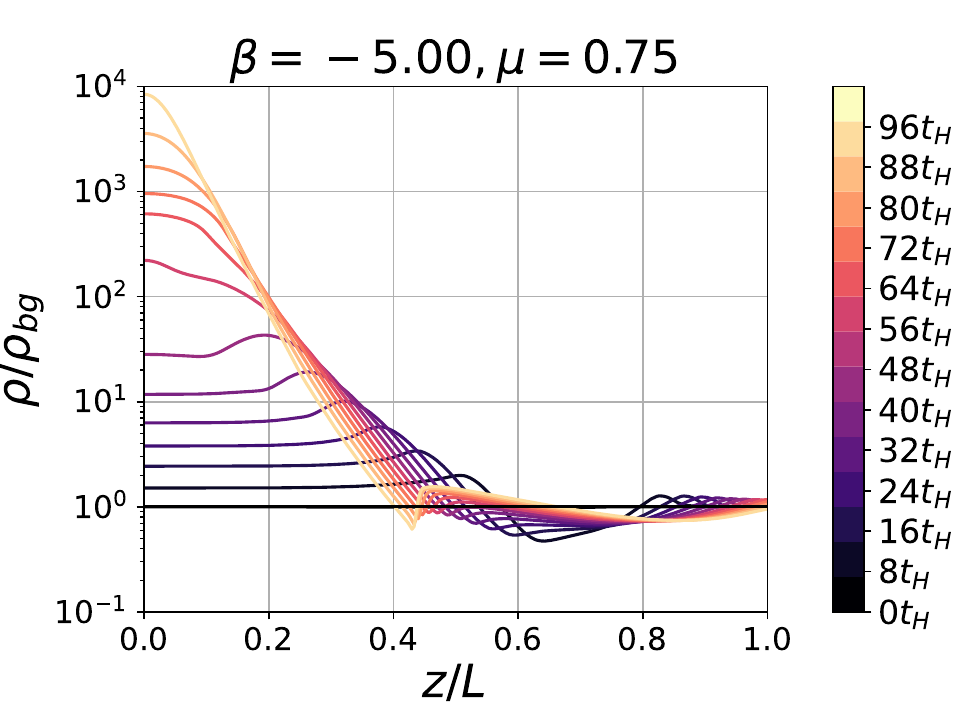}
    \end{minipage}
    \caption{The time evolution of $\rho/\rho_{bg}$ for $\beta =-5.00$ with $\mu = 0.70$ (left, sub-critical) and $\mu = 0.75$ (right, super-critical).
    }
    \label{delta_time}
    \end{figure}

We show the initial density profile at the critical amplitude of 
black hole formation for several values of
$\beta$ in Fig.~\ref{fig:initialdelta}. 
We should note that for negative $\beta$ case, $\rho/\rho_{bg}$ has a peak near 
$r/L\sim 0.3$, unlike in the positive $\beta$ case where $\rho/\rho_{bg}$ peaks at $r=0$. 
From Fig.~\ref{delta_time}, we can see that the growth of the second peak in its time evolution is smaller compared to the growth at $r=0$, and a high-density peak is realized at the center towards black hole formation as usual.
We also observe that for the case $\beta = -5$, a region with large gradients and underdensity is developed, in comparison with the other cases $\beta>-5$, which could explain the reason why such unexpected large fluctuations don't necessarily collapse forming a black hole. Nevertheless, further research is needed to clarify that aspect.

\subsection{PBH mass}\label{subsec:PBHmass}
\begin{figure}[htbp]
    \centering
    \begin{minipage}[h]{0.40\linewidth}
    \includegraphics[width = \linewidth]{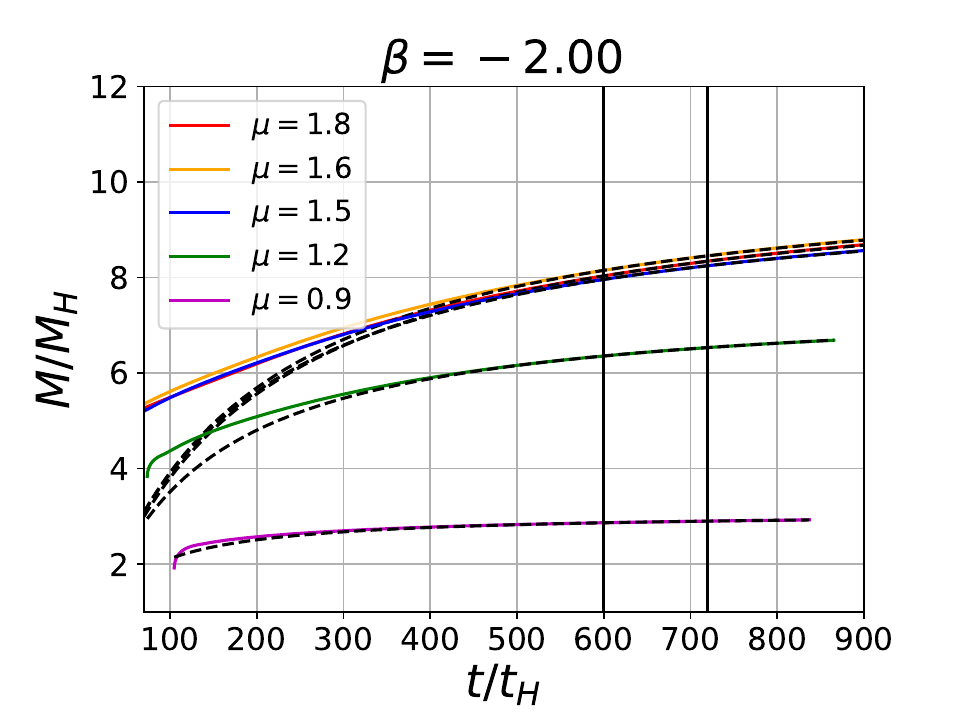}
    \end{minipage}
    \begin{minipage}[h]{0.40\linewidth}
    \includegraphics[width = \linewidth]{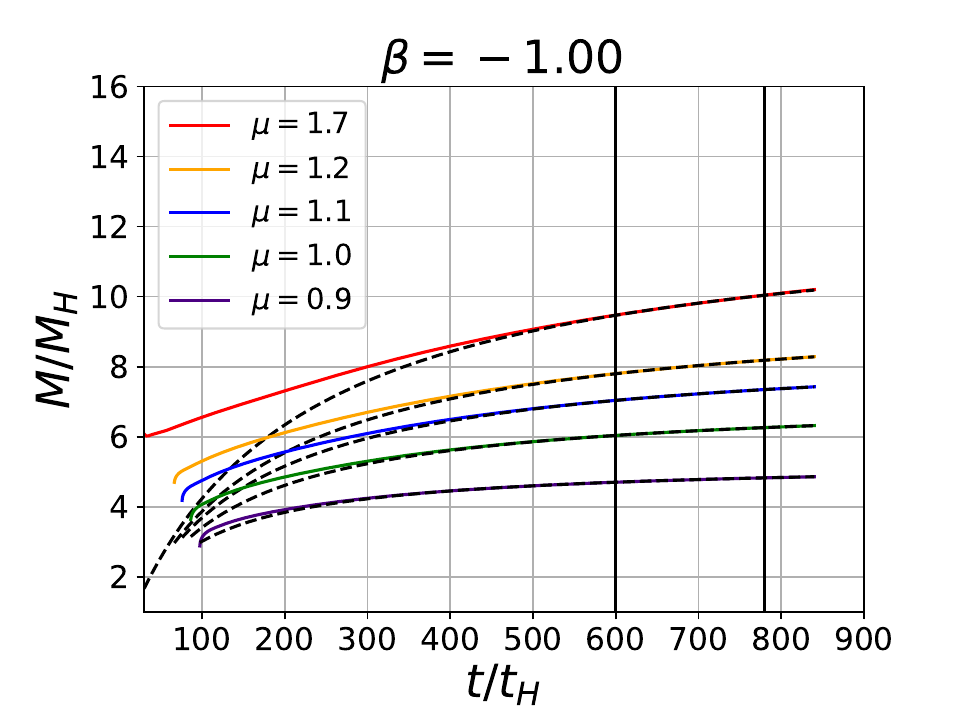}
    \end{minipage}
    \begin{minipage}[h]{0.40\linewidth}
        \includegraphics[width = \linewidth]{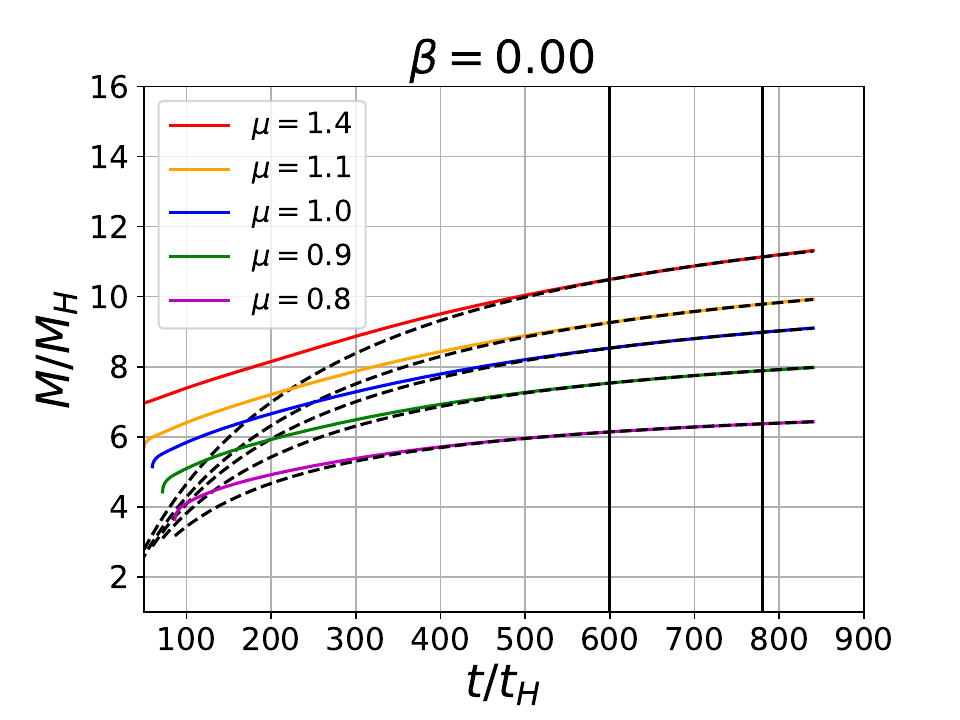}
    \end{minipage}
    \begin{minipage}[h]{0.40\linewidth}
    \includegraphics[width = \linewidth]{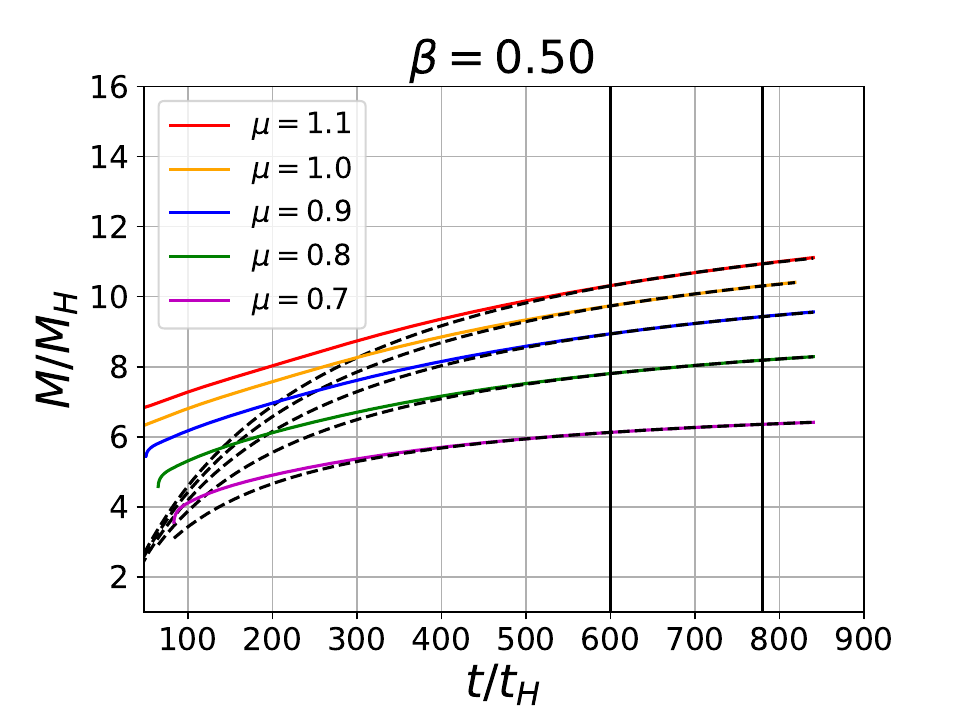}
    \end{minipage}
    \begin{minipage}[h]{0.40\linewidth}
    \includegraphics[width = \linewidth]{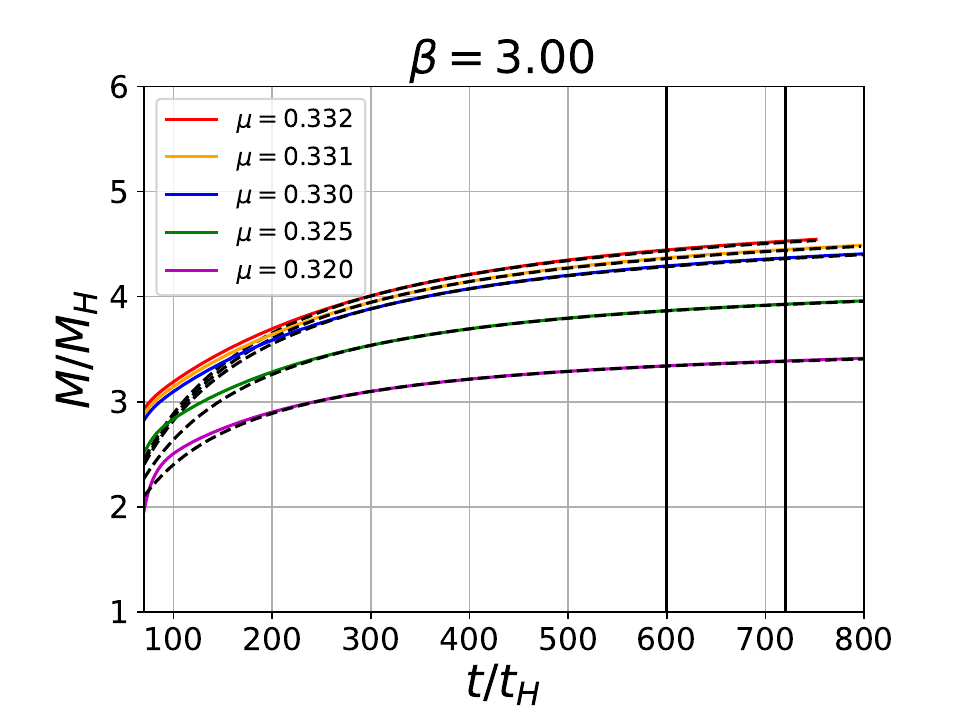}
    \end{minipage}
    \caption{The time evolution of PBH mass for $\beta = -2.00 $ (top-left), $\beta = -1.00$ (top-right), $0.00$ (middle-right), $0.50$ (middle-left), and $3.00$ (lower).
    Inside the black line is a fitting region, and the black dashed line is fitting. $M_H$ is the mass of the cosmological horizon at horizon reentry.}\label{fig:pbhmass}
\end{figure}

\begin{figure}[htbp]
    \centering
    \begin{minipage}[h]{0.40\linewidth}
    \includegraphics[width = \linewidth]{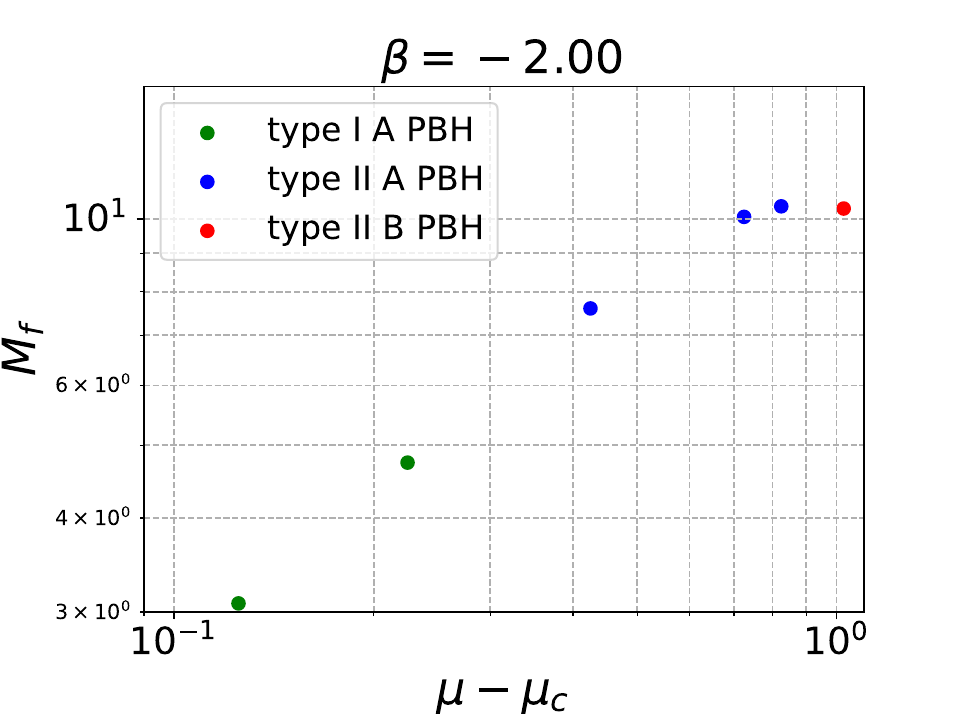}
    \end{minipage}
    \begin{minipage}[h]{0.40\linewidth}
    \includegraphics[width = \linewidth]{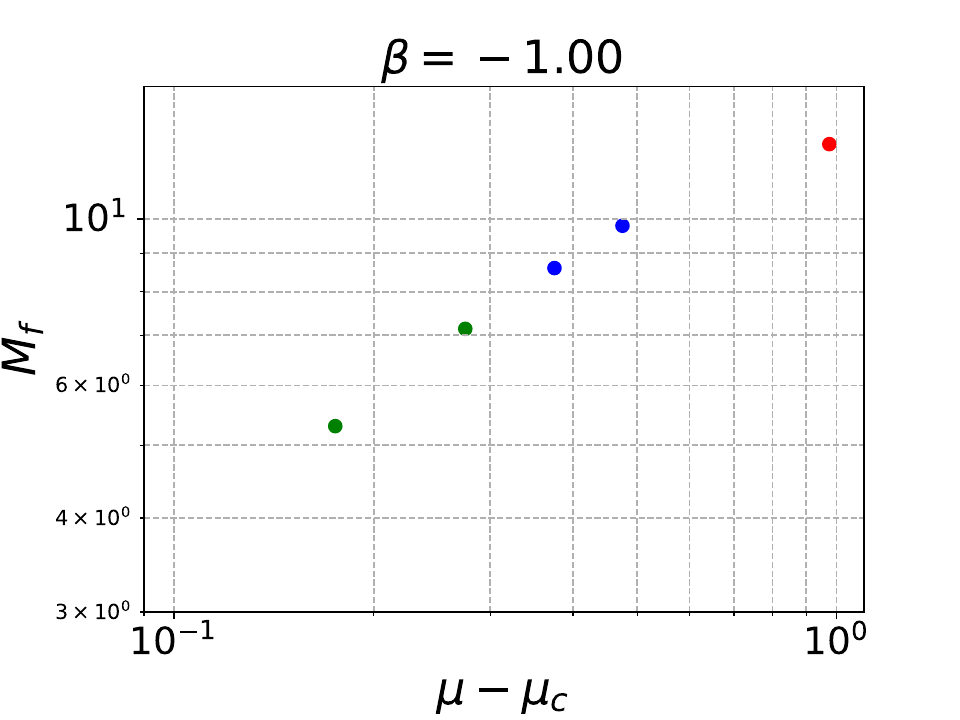}
    \end{minipage}
    \begin{minipage}[h]{0.40\linewidth}
        \includegraphics[width = \linewidth]{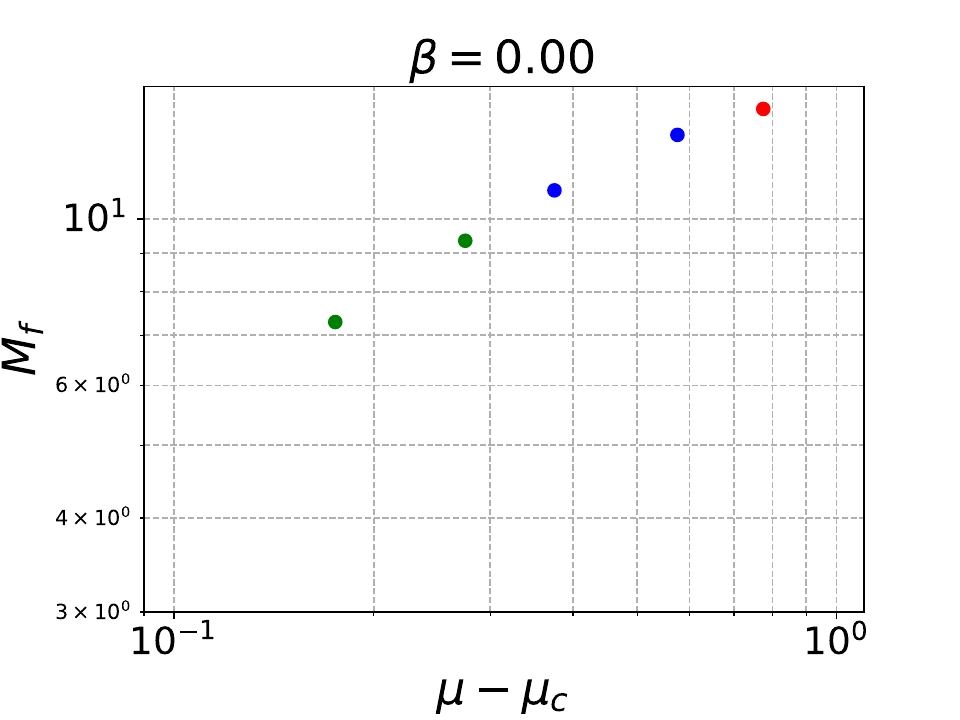}
    \end{minipage}
    \begin{minipage}[h]{0.40\linewidth}
    \includegraphics[width = \linewidth]{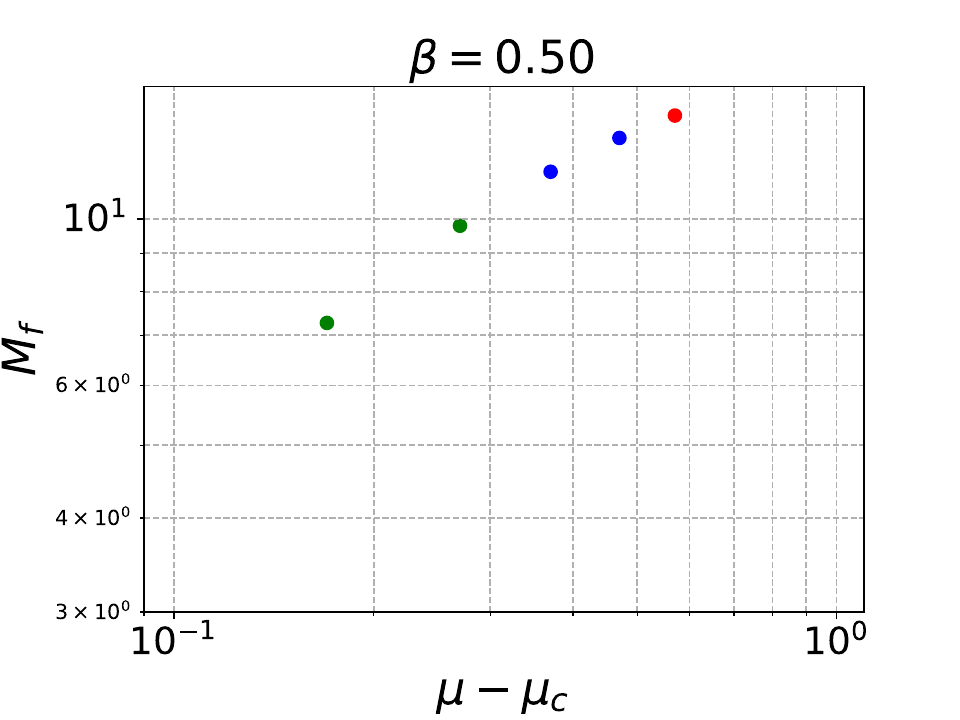}
    \end{minipage}
    \begin{minipage}[h]{0.40\linewidth}
    \includegraphics[width = \linewidth]{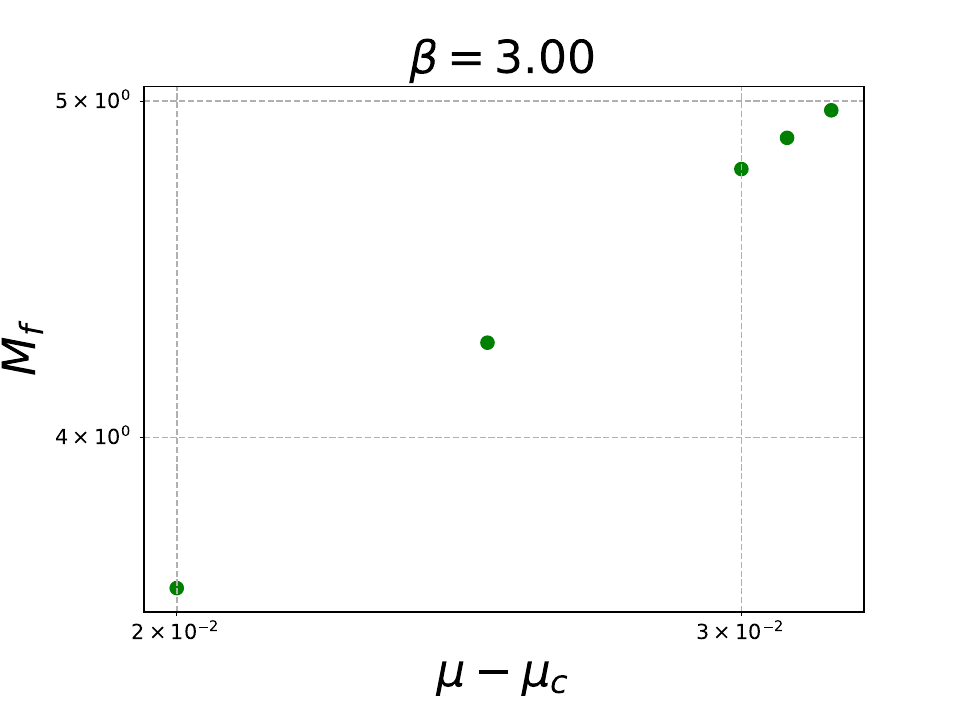}
    \end{minipage}
    \caption{The relation between the final black hole mass $M_{f}$ and $\mu-\mu_c$ in log scale, where $\mu_c$ is the threshold for black hole formation for different $\beta$ cases.
    The value of $\mu_c$ is determined as the average of the amplitudes of the blue and magenta dots in Fig.~\ref{fig:pbhmass}. The green dots represent type $\uI$ A PBHs, the blue dots represent type II A PBHs, and the red dot represents type II B PBH.}\label{fig:pbhmumuc}
\end{figure}

Let us consider the dependence of the PBH mass on $\mu$ and $\beta$. 
We follow the numerical evolution of the apparent horizon using an excision procedure 
and estimate the final mass fitting a line of the model known as the Novikov--Zeldovich or Bondi processes \cite{Zeldovich:1967lct,bondi} as is done in \cite{Deng:2016vzb,Escriva:2019nsa,Yoo:2021fxs,Uehara:2024yyp}.
In this model, the time evolution of mass can be expressed as follows
\begin{equation}
    \frac{dM}{dt} = 16\pi FM^2 \rho_b, 
    \label{eq:dmdt}
\end{equation}
where $\rho_b = \rho_{b,o}(t_0/t)^2$ is the background energy density, $t_0$ and $\rho_{b,0}$  being the initial time and the background energy density at that time, and $F$ is a constant that corresponds to the efficiency of accretion. 
Note that Eq.~\eqref{eq:dmdt} is not valid at the time of apparent 
horizon formation but valid at a later time when the accretion flow becomes stationary.
The integrated solution is 
\begin{equation}
    M(t) = \frac{1}{\frac{1}{M_f} + \frac{3F}{2t}}, 
    \label{eq:Mt}
\end{equation}
where $M_{f}$ is the final BH mass. 
We show the time evolution of the BH mass in Fig.~\ref{fig:pbhmass}.
By fitting the data in Fig.~\ref{fig:pbhmass} for sufficiently late times, between $t\sim 600 t_{H}$ and $t \sim 780 t_{H}$, we determine the value of $M_f$ and $F$, listed in Table.~\ref{tab:MfF}.
Fig.~\ref{fig:pbhmumuc} shows the relationship between the PBH final mass $M_{f}$ and the difference between the fluctuation amplitude and the threshold for PBH formation $\mu -\mu_c$ with $\mu_c$ being 
the value of $\mu$ indicated by the purple points in Fig.~\ref{fig:mucbeta}. 

\begin{table}[t]
    \centering
    \begin{minipage}{0.40\linewidth}
    \centering    
    \textbf{$\beta = -2.00$}\\
    \begin{tabular}{c|c|c|c}
    \hline
    \hline
     $\mu$ &  $M_{\rm f}[M_{H}]$& $F$&$M_{\rm f}/M_{\rm ini}$\\
     \hline
    0.90& 3.09 &5.00&4.82 \\
    1.20& 7.60& 5.10&5.93\\
    1.50& 10.06& 5.25&6.25\\
    1.60& 10.39& 5.30&6.04\\
    1.80& 10.32 & 5.52&6.18 
    \end{tabular}
    \end{minipage}
    \begin{minipage}{0.40\linewidth}
    \centering    
    \textbf{$\beta = -1.00$}\\
    \begin{tabular}{c|c|c|c}
    \hline
    \hline
     $\mu$ &  $M_{\rm f}[M_{H}]$& $F$&$M_{\rm f}/M_{\rm ini}$\\
     \hline
        0.90& 5.30& 4.76&5.52\\
        1.00& 7.14& 5.11&11.9\\
        1.10& 8.60& 5.16&12.23\\
        1.20& 9.79& 5.21&12.55\\
        1.70& 12.57& 5.22&12.57\\
    \end{tabular}
    \end{minipage}
    \begin{minipage}{0.40\linewidth}
    \centering
    \textbf{$\beta = 0.00$}\\
    \begin{tabular}{c|c|c|c}
    \hline
    \hline
     $\mu$ &  $M_{\rm f}[M_{H}]$& $F$&$M_{\rm f}/M_{\rm ini}$\\
     \hline
        0.80& 7.29& 5.12&11.95\\
        0.90& 9.35& 5.17&12.64\\
        1.00& 10.91& 5.11&12.68\\
        1.20& 12.93& 4.94&12.55\\
        1.40& 14.00& 4.79&12.17\\
    \end{tabular}
    \end{minipage}
    \begin{minipage}{0.40\linewidth}
    \centering
    \textbf{$\beta = 0.50$}\\
    \begin{tabular}{c|c|c|c}
    \hline
    \hline
     $\mu$ &  $M_{\rm f}[M_{H}]$& $F$&$M_{\rm f}/M_{\rm ini}$\\
     \hline
        0.70& 7.27& 5.13&12.32\\
        0.80& 9.79& 5.19&12.88\\
        0.90& 11.55& 5.06&12.69\\
        1.00& 12.81& 4.93&12.43\\
        1.10& 13.72& 4.83&12.14\\
    \end{tabular}
    \end{minipage}
    \begin{minipage}{0.40\linewidth}
    \centering
    \textbf{$\beta =3.00$}\\
    \begin{tabular}{c|c|c|c}
    \hline
    \hline
     $\mu$ &  $M_{\rm f}[M_{H}]$& $F$&$M_{\rm f}/M_{\rm ini}$\\
     \hline
        0.320& 3.62& 4.67&6.46\\
        0.325& 4.25& 4.80&7.32\\
        0.330& 4.78& 4.83&9.02\\
        0.331& 4.88& 4.83&9.76\\
        0.332& 4.97& 4.83&11.23\\
    \end{tabular}
    \end{minipage}
    \caption{
    The value of PBH final mass $M_{f}$, the efficiency of accretion $F$ and $M_f/M_{ini}$ where $M_{ini}$ is the initial PBH mass at the time when the apparent horizon is formed, for different cases of $\mu$ and $\beta$.}
    \label{tab:MfF}
\end{table}
We can notice from Table.~\ref{tab:MfF} and Fig.~\ref{fig:pbhmumuc} that,
near the threshold, the final PBH mass and $F$ increase as $\mu$ increases. The behavior of the PBH mass and $F$ are not trivial for the values of $\mu$ much larger than the threshold.
When comparing the final mass at the same $\mu-\mu_c$ for different $\beta$ values, we can notice that increasing non-Gaussianity $\beta$ results in a larger mass of the PBHs.
This can be explained by taking into account the role of pressure gradients, where large ones (characterized by large $q$) decrease the PBH mass, whereas, for small $q$, the gradients get smaller, making the accretion more effective and increase the PBH mass \cite{Escriva:2021pmf}.

Finally, from Fig.~\ref{fig:pbhmumuc}, we also observe that the PBH mass is a monotonic increasing function in terms of $\mu-\mu_c$ at least in the region near the threshold, which is relevant to observations. 
This differs from previous analytical expectations \cite{Kitajima:2021fpq}, where the behaviour of the PBH mass in the critical regime \cite{Niemeyer:1999ak} was expected to be
\begin{equation}
M = \mathcal{K} \,x^2_m(\mu) e^{2\zeta(\mu,r_m(\mu))}M_{k}(k_*)(\mu-\mu_c)^{\gamma},
\end{equation}
where $x_m=k_* r_m(\mu)$, $M_{k}(k_*)$ is the mass of the horizon associated to the wave-mode $k_*$, $\mathcal{K}$ is a profile dependent coefficient and $\gamma$ is a universal exponent only dependent on the equation of state of the perfect fluid \cite{choptuik,Evans:1994pj,Koike:1995jm,Niemeyer:1999ak,Musco:2012au}. 
For large $\beta$, the radius at the peak
of the compaction function decreases significantly with increasing $\mu$ (see Fig.~\ref{fig:rmmu}). 
Since the length scale $r_m$ is considered to be the major factor that determines the PBH mass,
this behaviour disrupts the monotonic increase in the mass function. 
Contrary to the expectation, our numerical results show an increasing monotonic behaviour with respect to $\mu$. A key implication of this finding is that the PBH mass function does not experience a divergence associated with the Jacobian term (when transforming the variable $\mu$ to $M$), which is linked to the non-monotonic behaviour of the PBH mass as suggested in \cite{Kitajima:2021fpq}. 
Nevertheless, notice that the divergent feature indicated by an analytic estimation 
was found to be slightly after the peak of the mass function \cite{Kitajima:2021fpq}, 
and the total PBH abundance is not significantly affected by the divergence of the mass distribution function.
It should also be noted that,
for $\beta=3$, the bubble channel is expected to dominate over the adiabatic one, according to \cite{bubble}.

\section{Summary}
\label{conclusion}
In this work, we have mainly investigated PBH formed from non-Gaussian and type II fluctuations through numerical simulations during the radiation-dominated era, 
where type $\uII$ fluctuations are characterized by the non-monotonicity of the areal radius $R(r)$. We have studied the initial shapes of $\zeta$, $\mathcal{C}$, the PBH formation dynamics, the classification of type A/B PBH and the PBH mass for different non-gaussian $\beta$ values.
We have found that the threshold for black hole formation and type A/B PBH classification depends on the non-Gaussianity $\beta$.
More specifically, 
the threshold values of the initial amplitude $\mu$ for black hole formation and type A/B PBH classification decrease with $\beta$
for $ \beta \geq -2$, and roughly saturate 
to a constant value in the range $-2 \leq \beta \lesssim -1$. For $\beta \gg 1$ we see that the threshold between type A/B matches with $\mu_{II}$.

Furthermore, when $\beta$ takes a significantly large negative value, specifically when $\beta \lesssim -4.0,$ the threshold for PBH formation lies in the type $\uII$ region.
This means that there are cases where type $\uII$ fluctuations do not form PBHs, which is referred to as "type $\uII$ no PBH". Then, the abundance of PBHs will be sourced, in principle, by fluctuations of type $\uII$ rather than type $\uI$, although a detailed calculation is beyond the scope of this paper.
Additionally, we found that the analytical estimations of the threshold for PBH formation 
proposed in \cite{deltacq}, especially regarding the $q$ approach described in section \ref{compaction}, well agree with our numerical results in the region of applicability for type I fluctuations.
Finally, we have found that the PBH mass increases monotonically with $\mu$ near the threshold for PBH formation, regardless of the value of $\beta$.
However, the behavior becomes non-trivial for the values of $\mu$ much larger than the threshold. Additionally, when comparing PBH masses for the same $\mu-\mu_c$ at different values of $\beta$, it has been found that a larger $\beta$ corresponds to a larger PBH mass.

\begin{acknowledgments}
While writing this paper, we learned that Ryoto Inui, Cristian Joana, Hayato Motohashi, Shi Pi, Yuichiro Tada, and Shuichiro Yokoyama were working on similar issues~\cite{Inui}. Although our research was conducted independently, we are grateful to them for pointing out the phenomenological importance of the cases of large negative $\beta$ values.
The authors are grateful to Jaume Garriga for useful discussions. A.E. acknowledges support from the JSPS Postdoctoral Fellowships for Research in Japan (Graduate School of Sciences, Nagoya University). D.S. is supported in part by JSPS KAKENHI Grant No. 24KJ1223. K.U. would like to take this opportunity to thank the “THERS Make New Standards Program for the Next Generation Researchers” supported by JST SPRING, Grant Number JPMJSP2125. C.Y. is supported in part by JSPS KAKENHI Grant Nos. 20H05850, 20H05853 and 24K07027.
\end{acknowledgments}

\clearpage
\appendix

\section{Modification of the shape due to the window function and its effects}
\label{apendix_window}

\begin{figure}[t]
    \centering
    \includegraphics[width = 0.4\linewidth]{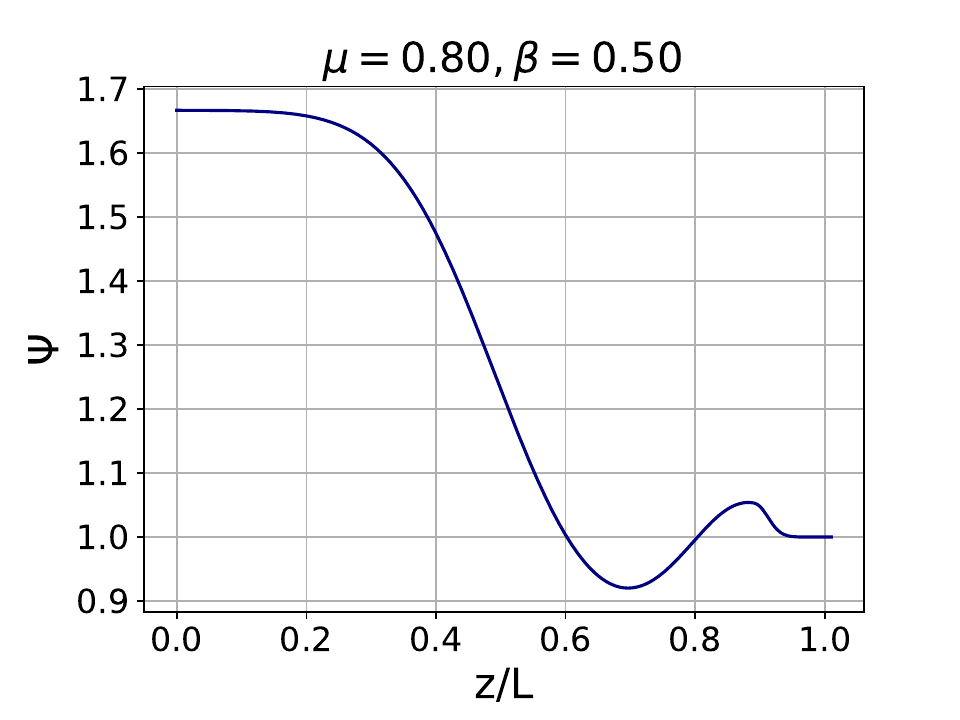}
    \includegraphics[width=0.4\linewidth]{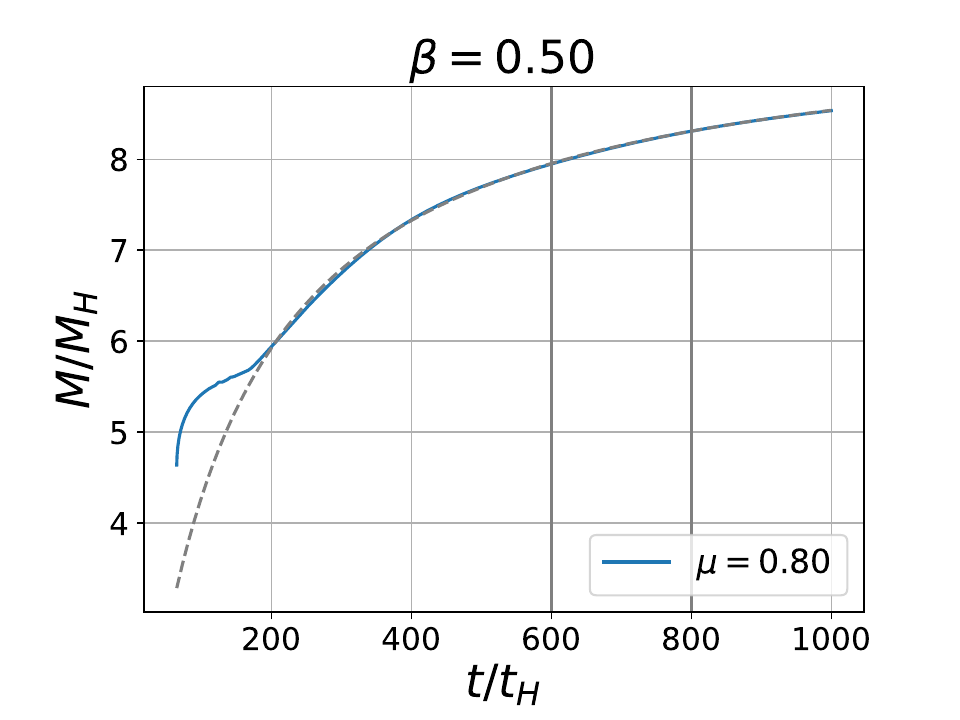}
    \caption{The initial profile of $\Psi$ such that it has two peaks (left-panel), and the time evolution of PBH mass for $\beta =0.50, \mu = 0.80.$ (right-panel). For this case, the final PBH mass is $9.60$, which is slightly different from the result found in Sec.\ref{subsec:PBHmass} with $9.79$ }
    \label{fig:extra_mas_behaviour}
\end{figure}

In this appendix, we give some insights and justification for the use
of the window function introduced in the section. The sinc profile is characterised by oscillations beyond the centre, which leads to a succession of peaks in the compaction function \cite{Atal_2020}. These oscillations are expected to have a very small effect on the threshold values since they are far away from the first peak of the compaction function, which leads to gravitational collapse. However, the effect of the PBH mass is more significant due to environmental effects (see \cite{Escriva:2023qnq} for a detailed study). This can be seen in Fig.~\ref{fig:extra_mas_behaviour} where changing the scale where the window function acts, we observe a steep increment of the PBH mass evolution. Therefore, our results can be slightly affected by the choice of the window function. Nevertheless, the dispersion of the shape can be large 
compared to the typical value in the far region, 
and therefore, we may have a significant uncertainty on the exact functional form of the shape \cite{1986ApJ...304...15B,Atal_2020}. For simplicity, in this work, we apply the window function to cut the additional details of the profile that may give to a non-trivial behaviour on the time evolution of the PBH mass, and we leave those aspects for future investigation.

\bibliographystyle{JHEP}
\bibliography{logzetav4}

\end{document}